\documentclass[aps,prd,nofootinbib,twocolumn,showpacs,superscriptaddress]{revtex4-1}
\newcommand{\ee}[1]{\!\times\!10^{#1}}

\pdfoutput=1
\usepackage {natbib}
\usepackage {graphicx}
\usepackage {hyperref}
\usepackage {color}
\usepackage{textcase}
\usepackage{amsmath, amsthm, amsfonts}
\usepackage{enumerate}
\usepackage{soul}
\usepackage[normalem]{ulem}

\begin{document}

\title{A comparison of methods for the detection of gravitational waves from unknown neutron stars}

\author{S. Walsh}
\affiliation{Max Planck Institute for Gravitational Physics (Albert Einstein Institut), Am M\"uhlenberg 1, 14476 Golm, Germany}
\affiliation{Max Planck Institute for Gravitational Physics (Albert Einstein Institut), Callinstrasse 38, 30167 Hannover, Germany}
\affiliation{Department of Physics, University of Wisconsin, Milwaukee, WI 53201, USA}
\author{M. Pitkin}
\affiliation{SUPA, School of Physics and Astronomy, University of Glasgow, Glasgow G12 8QQ, United Kingdom}
\author{M. Oliver}
\affiliation{Department de F\'isica-IAC3, Universitat de les Illes Balears and Institut d'Estudis Espacials de Catalunya,Cra. Valldemossa km. 7.5, E-07122 Palma de Mallorca, Spain}
\author{S. D'Antonio}
\affiliation{INFN, Sezione di Roma 2, Via della Ricerca Scientifica, 1, I-00133 Roma, Italy}
\author{V. Dergachev}
\affiliation{LIGO Laboratory, California Institute of Technology, MS 100-36, Pasadena, CA 91125, USA}
\author{A. Kr\'olak}
\affiliation{Institute of Mathematics, Polish Academy of Sciences, \'Sniadeckich 8, 00-956 Warszawa, Poland}
\author{P. Astone}
\affiliation{INFN, Sezione di Roma, P.le A. Moro, 2, I-00185 Roma, Italy}
\author{M. Bejger}
\affiliation{N. Copernicus Astronomical Center, Bartycka 18, 00-716, Warszawa, Poland}
\author{M. Di Giovanni}
\affiliation{INFN, Trento Institute for Fundamental Physics and Applications, I-38123 Povo, Trento, Italy}
\affiliation{Universit\`a di Trento, Dipartimento di Fisica, I-38123 Povo, Trento, Italy}
\author{O. Dorosh}
\affiliation{National Centre for Nuclear Research, 05-400 Otwock, \'Swierk, Poland}
\author{S. Frasca}
\affiliation{INFN, Sezione di Roma, P.le A. Moro, 2, I-00185 Roma, Italy}
\author{P. Leaci}
\affiliation{Universit\`a di Roma ``La Sapienza'', I-00185 Roma, Italy}
\affiliation{INFN, Sezione di Roma, P.le A. Moro, 2, I-00185 Roma, Italy}
\author{S. Mastrogiovanni}
\affiliation{Universit\`a di Roma ``La Sapienza'', I-00185 Roma, Italy}
\affiliation{INFN, Sezione di Roma, P.le A. Moro, 2, I-00185 Roma, Italy}
\author{A. Miller}
\affiliation{Universit\`a di Roma ``La Sapienza'', I-00185 Roma, Italy}
\affiliation{University of Florida, Gainesville, Florida 32611, USA}
\author{C. Palomba}
\affiliation{INFN, Sezione di Roma, P.le A. Moro, 2, I-00185 Roma, Italy}
\author{M. A. Papa}
\affiliation{Max Planck Institute for Gravitational Physics (Albert Einstein Institut), Am M\"uhlenberg 1, 14476 Golm, Germany}
\affiliation{Max Planck Institute for Gravitational Physics (Albert Einstein Institut), Callinstrasse 38, 30167 Hannover, Germany}
\affiliation{Department of Physics, University of Wisconsin, Milwaukee, WI 53201, USA}
\author{O. J. Piccinni}
\affiliation{Universit\`a di Roma ``La Sapienza'', I-00185 Roma, Italy}
\affiliation{INFN, Sezione di Roma, P.le A. Moro, 2, I-00185 Roma, Italy}
\author{K. Riles}
\affiliation{University of Michigan, 450 Church Street, Ann Arbor, Michigan 48109, USA}
\author{O. Sauter}
\affiliation{University of Michigan, 450 Church Street, Ann Arbor, Michigan 48109, USA}
\author{A. M. Sintes}
\affiliation{Department de F\'isica-IAC3, Universitat de les Illes Balears and Institut d'Estudis Espacials de Catalunya,Cra. Valldemossa km. 7.5, E-07122 Palma de Mallorca, Spain}

\begin{abstract}
Rapidly rotating neutron stars are promising sources of continuous gravitational wave radiation for the LIGO and Virgo interferometers. The majority of neutron stars in our galaxy have not been identified with electromagnetic observations. All-sky searches for isolated neutron stars offer the potential to detect gravitational waves from these unidentified sources. The parameter space of these blind all-sky searches, which also cover a large range of frequencies and frequency derivatives, presents a significant computational challenge. Different methods have been designed to perform these searches within acceptable computational limits. Here we describe the first benchmark in a project to compare the search methods currently available for the detection of unknown isolated neutron stars. We employ a mock data challenge to compare the ability of each search method to recover signals simulated assuming a standard signal model. We find similar performance among the short duration search methods, while the long duration search method achieves up to a factor of two higher sensitivity. We find the absence of second derivative frequency in the search parameter space does not degrade search sensivity for signals with physically plausible second derivative frequencies. We also report on the parameter estimation accuracy of each search method, and the stability of the sensitivity in frequency, frequency derivative and in the presence of detector noise.

\end{abstract}

\maketitle

\section{Introduction}

Continuous gravitational waves (CWs) from isolated neutron stars (NSs) are a potential source of detectable gravitational waves. CW radiation is emitted by rotating NSs with non-axisymmetric deformations. The signal is expected to be relatively stable over many years. While the amplitude of CW signals is expected to be small, the continuous nature of the signal allows us to integrate the signal over large time spans of data to distinguish it from noise.

Broad-band all-sky searches cover the whole sky over a broad range of frequency and frequency derivative in order to detect CW radiation from unknown NSs. All-sky searches in the initial Laser Interferometer Gravitational-Wave Observatory (LIGO) \cite{LIGO_Virgo_1,LIGO_Virgo_2} and Virgo \cite{LIGO_Virgo_3} data have so far not resulted in detection. Instead upper limits have been placed on the amplitude of CWs from isolated NSs \cite{EaH_S6, PF_2, FH_3, TD_1, SH_2}. The advanced detectors, which began operation in 2015, will eventually have a sensitivity to these weak signals over an order of magnitude more than that of the previous generation, with the largest gains at frequencies below 100\,Hz. 

The purpose of the study presented in this article is to examine and compare the efficiency of the methods that will be used to perform all-sky searches in data from the advanced LIGO and Virgo detectors. This comparison is made using a mock data challenge (MDC), for the standard CW signal model described in Sections \ref{sec:signal} and \ref{sec:MDC}. In a blind all-sky search, detectable CW signals may deviate from this model, for example if the NS glitches. For an accurate comparison of the all-sky search methods, further studies will be needed which include deviations from the standard CW model. The study presented here serves as a first benchmark for the search methods, assuming the signal consistently follows the model.

A brief overview of the search methods is presented in Section \ref{sec:methods}; the search parameters for the various searches are presented in \ref{sec:implementation}. Section \ref{sec:comparison} describes how the methods are compared. The results of the comparison are presented in Section \ref{sec:results}. \\

\section{The signal}
\label{sec:signal}

Gravitational waves (GWs) emitted from non-axisymmetric NSs are typically described by a signal model which remains relatively stable over years of observation \cite{Fstat}. The strain amplitude of the GW is proportional to the ellipticity, $\varepsilon$, defined as 
\begin{equation}
\varepsilon = \frac{|I_{xx} - I_{yy}|}{I_{zz}},
\end{equation}
where $I_{zz}$ is the principal moment of inertia of the star, and $I_{xx}$ and $I_{yy}$ are the moments of inertia about the axes. The strain amplitude of the GW at the detector, assuming a rigidly rotating triaxial body, is then given by 
\begin{equation}
h_0 = \frac{4\pi^2G}{c^4}\frac{I_{zz}f^2\varepsilon}{d},
\end{equation}
where $f$ is the frequency of the GW, $G$ is Newton's constant, $c$ is the speed of light, and $d$ is the distance to the NS. For a star steadily rotating around a principal axis of inertia, the frequency of the GW is at twice the rotational frequency of the NS. The frequency evolves over time as energy is lost due to various dissipation mechanisms, including GW emission. The first time derivative of the frequency, $\dot{f}$, is referred to as spindown.

The signal arriving at the detectors is modulated by the motion of the Earth around the Sun and by the rotation of the Earth. Therefore, the signal in the detector reference frame depends on the sky position $(\alpha,\delta)$ of the source. 

The signal model is described by eight parameters, four phase evolution parameters $(f_0,\dot{f},\alpha,\delta)$ and four amplitude parameters $(h_0,\iota,\psi,\phi_0)$, where $\iota$ is the inclination angle between the line of sight to the NS and its rotation axis, $\psi$ is the polarisation angle and $\phi_0$ is the initial phase of the signal at a reference time. 

In a blind all-sky search there is also the potential for the detection of signals produced by different source models (e.g.\ $r$-modes \cite{Owen_f2}). The ability of the all-sky search methods to recover such signals is not examined in this study. Here we assume the signal follows the model described above.\\

\section{Current methods}
\label{sec:methods}

The most sensitive search for CW signals is performed with a fully coherent integration over a large timespan of data. The computational power required for the integration increases rapidly with the observation time of the data. When searching for CW signals over a broad frequency and spindown range, and over the whole sky, a fully coherent search quickly becomes computationally unfeasible. \cite{VirgoCompCost,Brady}

This is the motivation for semi-coherent search methods. The data is split into shorter segments which are searched separately with a coherent method using a coarse grid in parameter space. The results of the coherent search in each segment are then combined incoherently on a finer search grid. For limited available computing power, these semi-coherent search methods achieve a higher sensitivity than could be achieved with a fully coherent search with a tractable coherence time \cite{Brady}. 

Some searches use segments which are short enough, on the order of 1800\,s, so that the signal remains within a single Fourier frequency bin in each segment. In this case, the power of the GW signal is extracted with a single Fourier transform in each segment. Other searches use longer segment times, hours to days, to increase the SNR of the signal in each segment. In this case, the coherent integration uses the more computationally demanding $\mathcal{F}$-statistic \cite{Fstat} to take into account signal modulations. Each search method uses a different variable to quantify SNR, so the numeric values of the SNR thresholds used by each search can not be directly compared. 

The sensitivity of the semi-coherent searches is improved by taking a hierarchical approach. After the semi-coherent all-sky search, candidates are selected with a threshold which is lower than needed to claim a detection. A refined, more sensitive search is then performed in the parameter space surrounding each candidate. In principle, the significance of recovered candidates increases in the presence of signal, but not if an original candidate is due to a random noise outlier.

The refinement stages are designed such that any signal passing the first stage has a high probability of being recovered at each following stage. Therefore, the threshold at the first stage ultimately determines the sensitivity of the search.

The deepest searches are performed by the Einstein@Home pipeline, which benefits from the large computing power provided by the Einstein@Home project (Section \ref{sec:EaH_method}). Einstein@Home searches take many months before the presence of signal can be confirmed or excluded. There are also quick-look search pipelines which have a much shorter turnaround time. Each all-sky search makes different tradeoffs in the sensitivity vs.\ robustness against deviations from the assumed phase models.  In the following we provide a brief overview of the search procedure employed by each pipeline, and the distinguishing characteristics of each method. \\

\subsection{Powerflux}
\label{sec:PF_method}

The Powerflux method is described in \cite{SH_PF_1,PF_2,PF_3}. This search uses 1800-s Hann-windowed short Fourier Transforms (SFTs), with an effective coherence length on the order of 900 s. The power from each SFT is recorded along the track corresponding to each point in parameter space in the time-frequency plane, accounting for Doppler shift and spindown. The power is then weighted, according to noise and detector antenna pattern, to reduce outliers from noise artifacts and maximise the signal-to-noise ratio.

The dataset is partitioned into $\sim 1$ month stretches, and the sum of the weighted power along the track is produced independently for any contiguous combination of these stretches. High-SNR candidates are identified based on their persistence across contiguous stretches of data.

The candidates are then confirmed as signal or rejected as noise with four additional search stages around each candidate. The parameter space refinement increases with each stage, and the last three stages use the Loosely Coherent detection pipeline \cite{PF_LC_1}. In addition to searching over the four-dimensional parameter space $(f,\dot{f},\alpha,\delta)$, Powerflux also searches over polarisation angle $\psi$. \\ 

\subsection{Sky Hough}
\label{sec:SH_method}

The Sky Hough method is described in \cite{SH_PF_1,SH_2}. For this search, 1800-s SFTs are digitised by setting a threshold of 1.6 on their normalised power. Thereby, each SFT is replaced by a collection of zeros and ones called a peak-gram. The Hough number count is the weighted sum of the peak-grams along the track corresponding to each point in parameter space in the time-frequency plane, accounting for Doppler shift and spindown. This sum is weighted based on the detector antenna pattern and the noise level, to suppress outliers from detector artifacts. Candidates are selected based on the deviation of the weighted number count from its value in Gaussian noise. 

The data is split into two sets containing an equal number of SFTs. The search parameter space is split into sub regions in frequency and sky location. For every region the search returns a toplist of the most significant candidates for both datasets. These candidates are required to pass a significance threshold, and a $\chi^2$ test is applied to eliminate candidates coming from detector artifacts. Candidates which are not within a coincidence window of each other in both datasets are discarded.

A clustering algorithm is then applied to coincident candidates. The most significant cluster candidate per 0.1\,Hz is chosen based on its distance to all other candidates in the toplist, weighted by some significance. Passing candidates are confirmed or rejected with a refinement stage, which covers a reduced parameter space around each candidate with higher resolution in spindown and sky.\\

\subsection{Time domain $\mathcal{F}$-Statistic}
\label{sec:TD_method}

The Time domain $\mathcal{F}$-statistic search method uses the algorithms and pipeline described in \cite{TD_1,TD_sky}. This analysis uses fast Fourier Transformed data segments of two sidereal days each, split into bands of 0.25\,Hz. A coherent search is performed in each segment, and candidates with an $\mathcal{F}$-statistic above threshold are recovered. Recovered candidates around known detector artifacts are vetoed, as are those with similar profiles to stationary noise lines and those close to the polar caps in equatorial coordinates.

The method then searches for coincidences among candidates in each two-day segment. Candidate frequencies are converted to a common reference time, using the candidate spindown. Coincident candidates are counted, within a coincidence window large enough to account for Doppler modulation. The coincidence with the highest multiplicity is considered the most significant candidate, which is then selected or rejected based on a threshold on its false alarm probability. \\

\subsection{Einstein@Home}
\label{sec:EaH_method}

Einstein@Home is a volunteer-driven distributed computing project where members of the public donate their idle computing power to the search for GWs \cite{EaH_url}.  The donated computing power allows for broader and more sensitive searches for CWs.
The Einstein@Home search is described in \cite{EaH_S6}. The search begins with 1800-s SFTs. SFT bins which overlap with known detector artifacts are cleaned by replacing them with Gaussian noise. For the coherent analysis, the SFTs are combined into segments of a few days. The $\mathcal{F}$-statistic is computed for each segment and for each parameter space point on a coarse grid. An average $2\mathcal{F}$-statistic is then calculated by summing the $2\mathcal{F}$ values at each segment at the appropriate coarse grid point to approximate the $2\mathcal{F}$ at a given fine grid point \cite{GCT_method}. The logBSGL statistic, described in \cite{BSGL}, is calculated for each point. This is derived from the $2\mathcal{F}$-statistic, and suppresses detector artifacts appearing in one detector.

The search parameter space is split into regions in frequency and sky. The search is performed for each region, and a toplist of the candidates ranked by logBSGL is returned. Candidates from the toplist are selected for further study based on their $2\mathcal{F}$ value. The threshold applied depends on the total number of candidates above this threshold. There must be few enough that they can all be studied further with a refined search. These candidates are expected to be predominantly from Gaussian noise. 

The search proceeds with multiple stages of refinement, described in \cite{EaH_S6_FU}, to confirm or reject the presence of a signal.

\subsection{Frequency Hough}
\label{sec:FH_method}
The Frequency Hough method is described in \cite{FH_1,FH_2,FH_3}.  The analysis uses time-domain cleaned SFTs with a timespan which depends on the frequency band of the search, and is chosen so that the signal remains within a single Fourier frequency bin. 
A time-frequency map (peakmap) is constructed from the database by selecting the most significant local maxima on the square root of the equalised power\footnote{The equalized power is obtained by dividing the periodogram squared by an auto-regressive estimation of the average power. The result is a quantity that typically has a value close to unity except in correspondence of narrow spectral lines, where it takes values larger than one.}, called peaks, over a threshold of $\sqrt{2.5} \simeq 1.58$.

The peakmap is cleaned by removing peaks corresponding to lines at a fixed frequency and to wandering lines. Because of time constraints, this cleaning is not applied in the MDC. 

In the Frequency Hough step, the points of the corrected peakmap are mapped onto the signal frequency/spindown plane for every sky position. The parameter space (sky position, frequency, spindown) is suitably discretised. In particular, the frequency resolution is increased by a factor of 10 with respect to the `natural' choice (given by the inverse of the SFT duration). The adaptive procedure on the Hough transform, that would allow to take into account noise non-stationarity, has not been used in this analysis.

A given number of the most significant candidates are selected at each sky position and in each 1-Hz interval. This avoids being blinded by particularly disturbed frequency bands. For each candidate a search, refined in spin-down and sky resolution, is performed around the candidate parameters. The refined candidates are clustered and their coincidences with the candidates of another dataset are computed. The coincident candidates are then ranked by significance and the most significant candidates are subject to a refinement stage.\\

\section{The mock data challenge}
\label{sec:MDC}
The aim of the MDC is to empirically compare the performance of current all-sky search methods when searching for a standard CW signal from an isolated NS. This is done by simulating the detector response to CW signals in data from the S6 LIGO science run \cite{S6data}, with software injections at a range of frequencies. Each of the pipelines described in Section \ref{sec:methods} then performs a search over the data to assess their ability to recover this signal. \\

\subsection{The data}
\label{sec:the_data}

The MDC search is performed over data from the LIGO S6 science run, in which simulated CW signals are injected. Real LIGO data is used to assess the performance of search methods in the presence of detector artifacts. The software injections are generated with {\tt lalapps\_sw\_inj\_frames} in the LALSuite software package \cite{Lalsuite}. There are 3110 injections in total. In general, the SNR of the injections was drawn randomly from a uniform distribution between 0 and 150 for a coherent single detector analysis in S6 H1 data (15 months). 50 of the injections have a coherent SNR between 1000 and 2000.

Between 40--1550\,Hz the injections are placed at 0.5\,Hz intervals, while 90 further injections between 1550--2000\,Hz are placed at 5\,Hz intervals. The sky position is isotropically distributed over the sky sphere. The spindown is randomly drawn from a uniform distribution in log space between $-1\ee{-9}$ and $-1\ee{-18}$\,Hz/s for $95\%$ of injections and between $1\ee{-18}$ and $1\ee{-13}$\,Hz/s for $5\%$ of injections\footnote{Some injections have been given spinups as real signals can have observed (rather than necessarily intrinsic) spinups due to, e.g.\ large proper motions for nearby sources, or accelerations in the centers of globular clusters}. A braking index between $n=5$ and $n=7$ ($n$ defined implicitly in Equation \ref{eqn:braking_index}) is applied to 25\% of the pulsars with spindown. The braking index of $n=5$ is for a NS which loses all its rotational energy through emission from a constant mass quadrapole, while $n=7$ is for saturated $r$-mode emission \cite{Wette_f2,Cristiano_f2,Owen_f2}. From the braking index, and the assigned frequency and frequency derivative, the second and third frequency derivative are assigned via the equation 

\begin{equation}
\dot{f} = -Kf^n,\:\ddot{f} = \frac{n\dot{f}^2}{f},\:\dddot{f} = \frac{n\dot{f}}{f}\left(2\ddot{f}-\frac{\dot{f}^2}{f}\right).
\label{eqn:braking_index}
\end{equation}
The torque function, $K$, is described in \cite{AllenK}. The nuisance parameters of the NS, $\psi$, $\phi_0$ and $\cos \iota$, are randomly drawn from uniform distributions with the ranges [$-\pi/4$, $\pi/4$], [0, $2\pi$] and [$-1$, 1] respectively.

\subsection{The MDC search}

It is not feasible for all pipelines to perform a full all-sky search for the MDC, as limited computational resources must be reserved for searches for real signals. Instead, the search is performed over a reduced volume in parameter space around each injection. This MDC search volume is chosen to be small enough that even the most computationally expensive pipeline can participate in the MDC, and large enough so that the MDC results will accurately represent the result of a full all-sky search.

Each injection is placed roughly at the center of an MDC search volume of $0.1$\,Hz, max[$2\ee{-9}$\,Hz/s, $\pm 3 \times$ true spindown/spinup] and a region in sky with a radius of 30 degrees $\times$ min[200\,Hz/injection frequency, 1]. The pipelines are provided with the start point and width of the bands in frequency and frequency derivative, and the center and radius of the sky patch. 

The specifics of the MDC search performed by each pipeline are described in Section \ref{sec:implementation}. After the analysis, each pipeline provides a list of candidates considered to be detections, with the candidate frequency, spindown, right ascension, declination and reference time. 

The MDC is split into two stages, with each stage covering half of the signals. In the first stage the signal parameters are known. Some search methods use this stage to improve their candidate selection and refinement. In the second stage of the MDC, the signal parameters are unknown, only the boundaries of the search region for each injection are provided. These are referred to as blind injections. This stage is used to verify the results from the search over known injections. Once verified, the results from all injections are combined for the final comparison between pipelines.\\

\subsection{Defining detections}
\label{sec:MDC_define_detection}
The search parameters and selection criteria used in all-sky searches depend on the parameter space to be covered. For the MDC, each search is required to use search parameters, significance thresholds and selection criteria that would be used in a search over the complete parameter space in frequency, spindown and sky. 

In an all-sky search, all candidates from the initial search would pass through the refinement stages. The number of candidates which can be followed up is limited, and different for each method. The detection threshold applied in the MDC must be shown to result in a manageable number of candidates from noise in a full all-sky search for each pipeline. Therefore, each pipeline must establish the number of false alarms expected in a full all-sky search with the thresholds applied in the MDC.

As mentioned in Section \ref{sec:methods}, the selection criteria applied after the initial search ultimately determine the ability to recover signals, as the refinement stages primarily exclude noise. To avoid excessive computing cost, we determine which candidates would result in a detection without requiring them to pass through all refinement stages. For the Sky Hough and Time domain $\mathcal{F}$-statistic searches, no false alarms from random noise outliers are expected among the candidates selected within the MDC search volume. Therefore, all candidates are considered detections. The Powerflux candidates are passed through the first refinement stage, after which no false alarms from random noise outliers are expected in the MDC, so all candidates after the first refinement stage are considered detections.

The Einstein@Home search does expect false alarms from random noise outliers after the initial search in the MDC, as it is set up to refine many more candidates than the other searches. The Frequency Hough search also expects false alarms because it searches the whole parameter space, not just the reduced MDC search volume. In a real search, all of these candidates would be passed through the refinement stages, and candidates from signals would result in detections. In the MDC, we apply a threshold on the distance in parameter space between the signal and the recovered candidates in order to identify the candidates from the initial search which would result in a detection after the refinement stages.

For all pipelines, the refined searches are performed for a few MDC candidates that are close to the selection threshold to demonstrate that signals within this distance threshold are retained after the refinement. We also look at the distance in each dimension of parameter space between the signal and the recovered candidates to check for outliers. \\

\section{Comparison of methods}
\label{sec:comparison}

\subsection{Detection efficiency}

Here we are primarily concerned with the ability of different searches to recover the CW signals. The detection efficiency is the fraction of signals which are considered detected, and it is the benchmark that we will use to compare pipeline performance. The detection efficiency is measured as a function of signal strength, $h_0$, expressed by the sensitivity depth $\sqrt{S_{h}}/h_0\, (1/\sqrt{\mathrm{Hz}})$, where $S_h$ is the harmonic sum over both detectors of the power spectral density of the data, at the frequency of the signal. At fixed detection efficiency, a more sensitive search will detect a weaker signal, i.e. it will have a higher sensitivity depth. 

Some injections overlap with known detector artifacts. We examine the detection efficiency separately for these signals to assess the performance of the methods in noisy data. We also separate the detection efficiency by frequency, as S6 data contains fewer instrumental artifacts at frequencies greater than 400\,Hz. The detection efficiency is also assessed for signals with positive frequency derivative, and those with non-zero second order spindown. \\

\subsection{Parameter estimation}
\label{sec:parameter_estimation}

In a broad parameter space search, the reduction in parameter uncertainty is achieved through the refinement stages that follow the original search. In the MDC the refinement stages are not systematically carried through, as explained in Section \ref{sec:MDC_define_detection}. We examine the parameter uncertainty of detection candidates at the final MDC stage and discuss how each method plans to reduce this uncertainty to the level required for a confident detection. In most cases, this automatically yields good parameters estimation (see e.g. \cite{Schaltev}). \\

\subsection{Computational cost}
\label{sec:computational_cost}

Each of the methods has made different compromises on sensitivity to develop an all-sky search using available computational resources. 

The MDC is performed on different CPUs for each method, and some chose to cover a larger parameter space than required. Therefore, an accurate comparison of the computing time of each method based on the MDC results is not possible. Instead, each method provides an estimate of how much computing power is required to perform a realistic all-sky search over the first four months of advanced LIGO data (see Table \ref{tab:computing_cost}). The computing cost for each method is provided in MSU (million standard units), where one SU is one core-hour on a standard core. The standard core used here is an Intel Xeon E5-2670 CPU.

The estimates in Table \ref{tab:computing_cost} are for the four month observing time of the first advanced LIGO data, which is different to the 9 or 15 month observing times used in the MDC. The estimates are also for searches over different ranges in frequency and spindown for each pipeline. It gives a rough idea of the difference in computing resources actually used by each pipeline. These estimates also do not take into account tuning cost and postprocessing costs. By design Einstein@Home is the most computationally intensive, as it is intended to run on the Einstein@Home grid. \\ 

\begin{table}[h!]
\begin{tabular}{|lc|}
\hline
Pipeline & Expected runtime of O1 search\\
\hline
Powerflux& 6.8 MSU\\
Time domain $\mathcal{F}$-statistic & 1.6 MSU\\
Frequency Hough & 0.9 MSU\\
Sky Hough & 0.9 MSU\\
\hline
Einstein@Home & 100 - 170 MSU \\
\hline
\end{tabular}
\caption{Expected computational costs of searches using the first four months of advanced LIGO data with each search pipeline. These estimates are for a different data observing time from that of the MDC, and do not cover the same parameter space as each other or the MDC. The Einstein@Home searches uses the computing resources of the Einstein@Home project and is designed to run for 6 - 10 months in the Einstein@Home grid. }
\label{tab:computing_cost}
\end{table}

\section{Implementation}
\label{sec:implementation}

In this section we detail the specific search parameters used by each pipeline for the MDC, and explain why the thresholds chosen here are representative of the values in an all-sky search.

The search grid parameters and thresholds applied here will vary in future searches, depending on the observation length of available data, how well behaved the data is, the parameter spacing being covered by the search, and other factors. Any changes to these parameters will affect the detection efficiency, and the variations will be different for each search method. When presenting the MDC results we consider only statistical uncertainties on the measured efficiency.  One should keep in mind that these results are specific to the search implementation presented here. 

Instances where the searches were not optimal in the MDC are highlighted. In some cases, predictions for how the sensitivity will change in future searches are included.\\

\subsection{Powerflux}

The Powerflux MDC search uses the same search parameters as the Powerflux all-sky search over S6 data described in \cite{PF_3}, these are summarised in Table \ref{tab:PF_setup}. The S6 search was not tuned for frequencies below 400\,Hz, therefore, a loss in performance at low frequencies is expected in the MDC. The tuning for searches in advanced detector data will include the low frequency range, reducing or removing this loss in performance.

The search uses all of the MDC data. The sky grid is isotropic on the celestial sphere, with the angular spacing between grid points given by the formula

\begin{equation}
\frac{4500}{T_{coh}\times(f_0+f_1)\times0.5\times\mathrm{sky\; refinement}},
\label{eq:PF_sky}
\end{equation}
where $f_0$ and $f_1$ are the start and end frequencies of each 0.25\,Hz band, and $T_{coh}$ is the coherent segment length.

Every candidate with an SNR greater than 5 is selected. In order to pass to the first refinement stage, candidates are required to appear in at least six contiguous stretches of data. The number of false alarms in a real search is expected to be dominated by instrumental artifacts, and so is difficult to predict. However, since the MDC search uses the same parameters as a previous all-sky search over S6 data, we know the number of false alarms is manageable \cite{PF_3}. 

Candidates surviving the first refinement stage would normally be passed through further stages, with less than $1\%$ false dismissal in the subsequent stages. For the parameter space covered by the MDC, the false alarm rate after the first refinement stage is expected to be negligible. For the MDC, a signal is considered detected if a candidate survives stage one, as all MDC candidates at this stage are expected to be from signals. This choice is justified by performing all refinement stages for the weakest candidates and demonstrating that they are recovered with high significance close to the signal.

\begin{table}[h!]
\begin{tabular}{|c|cc|}
\hline
Stage & 0 & 1 \\
\hline
$T_{coh}$ (s) & 900  & 900  \\
$\delta f$ (Hz) & $2.78\ee{-4}$ & $6.95\ee{-6}$ \\
$\delta \dot{f}$ (Hz/s) & $2\ee{-10}$ & $1\ee{-10}$\\
Sky refinement (rad) & 1 & 0.25\\
Phase coherence & NA & $\pi/2$ \\
\hline
\end{tabular}
\caption{Powerflux MDC search parameters. The sky resolution is given by the sky refinement as shown in Equation \ref{eq:PF_sky}.}
\label{tab:PF_setup}
\end{table}

\subsection{Sky Hough}

The Sky Hough MDC search uses a similar search grid to a previous all-sky search over LIGO data \cite{SH_2}, given in Table \ref{tab:SH_setup}.
The search uses all of the MDC data. The equatorial spacing of the sky grid points (in radians) is given by the formula

\begin{equation}
\frac{10^4\; \delta f}{f\times \mathrm{Pixel\; factor}}.
\label{eq:SH_sky}
\end{equation}

The post-processing procedure has been updated significantly since the previous search, as described in \cite{SH_new} and briefly in Section \ref{sec:SH_method}. The most significant cluster candidate is required to have SNR $\ge 4.5$, and pass the $\chi^2$ veto described in \cite{SH_2}. All surviving MDC candidates are considered detections. 

By selecting only the most significant cluster per frequency band, there is an upper bound on the number of candidates from the all-sky search to be followed up. Therefore, the number of surviving candidates would not be unreasonable in an all-sky search. \\

\begin{table}[h!]
\begin{tabular}{|c|c|}
\hline
Tcoh (s) & 1800\\
$\delta f$ (Hz) & $5.55\ee{-4}$ \\
$\delta \dot{f}$ (Hz/s) & $1.37\ee{-11}$ \\
pixel factor & 2 \\
\hline
\end{tabular}
\caption{Sky Hough MDC search parameters. The sky resolution is determined by the pixel factor as shown in Eq \ref{eq:SH_sky}.}
\label{tab:SH_setup}
\end{table}

\subsection{Time domain $\mathcal{F}$-statistic}

The Time domain $\mathcal{F}$-statistic search divides all of the MDC data into two-day segments, and uses 117 segments for the analysis based on the goodness of data \cite{TD_1}. The construction of the 4-dimensional search grid is described in \cite{TD_sky}, to achieve a minimal match of $\sqrt{3}/2$ using the smallest number of grid points with a frequency spacing of $5.79\ee{-6}$\,Hz. 

Candidates with $\mathcal{F} > 10.5$ (corresponding to an SNR of 4.1) in each segment are selected. The bandwidth of each segment is 0.25\,Hz, but only candidates in the 0.1\,Hz band defined by the MDC search are selected.

After counting coincidences across two-day segments, the most significant candidate per 0.1\,Hz band is considered a detection if it is coincident in at least 60 segments. With a simplified estimation of the false alarm rate, this corresponds with a false alarm probability of less than 0.1\% per 1\,Hz band. 

With only one candidate per 0.25\,Hz band and an additional threshold with 0.1\% false alarm probability, the number of false alarms will not become unmanageable in an all-sky search, and no false alarms are expected in the reduced parameter space covered by the MDC.\\

\begin{table}[h!]
\begin{tabular}{|c|c|}
\hline
Tcoh (h) & 48 \\
$\delta f$ (Hz) & $5.79\ee{-6}$ \\
Minimal Mismatch & $\sqrt{3}/2$ \\
\hline
\end{tabular}
\caption{Time domain $\mathcal{F}$-statistic search parameters. The minimal mismatch is used to construct the 4-dimensional search grid as described in \cite{TD_sky}.}
\label{tab:TD_setup}
\end{table}

\subsection{Einstein@Home}
\label{sec:EaH_imp}

To cover the available frequency band of 40 to 2000\,Hz, the Einstein@Home search uses three separate search configurations for 40 to 500\,Hz, 500 to 1000\,Hz and 1000 to 2000\,Hz. 
The search configurations are given in Table \ref{tab:EaH_setup}.

The searches use nine months of the MDC data. The sky grid is hexagonal and uniform on the ecliptic plane, with the distance between grid points given by
\begin{equation}
\frac{\sqrt{\mathrm{sky\; factor}}}{\pi \tau_E f},
\label{eq:EaH_sky}
\end{equation}
where $\tau_E$ is the radius of the Earth divided by the speed of light. 
The sky grid points are then projected to equatorial coordinates for the search.

The $2\mathcal{F}$ thresholds given in Table \ref{tab:EaH_setup} are chosen to result in 35 million false alarms in Gaussian noise in an all-sky search for each frequency band. The false alarm rate in Gaussian noise is estimated as described in \cite{EaH_S6}. 

In an all-sky search, selected candidates are passed through the refinement stages. In the MDC, the parameter space around the candidates in the first refinement stage is used to determine which candidates from the initial stage would result in a detection, see Section \ref{sec:MDC_define_detection}. 

For the 40 to 500\,Hz search, the first refinement stage searches $\delta f \pm 1.9\ee{-4}$\,Hz, $\delta \dot{f} \pm 3.46\ee{-11}$\,Hz/s and a sky patch with a radius of 1.2 initial-search sky grid bins around the selected candidate. Therefore, candidates within this region of the signal are considered detections. The 500 to 1000\,Hz search uses a larger parameter space for the first refinement stage, candidates within $\delta f < 6\ee{-4}$\,Hz, $\delta \dot{f} < 1.8\ee{-10}$\,Hz/s and 3.2 sky grid bins of the signal are considered detections. The distance threshold is the same for the 1000 to 2000\,Hz search, except that the candidate must be within 2.4 sky bins. In each case, $\sim 90\%$ of signals are expected to have a candidate within this region around the signal parameters.

\begin{table}[h!]
\begin{tabular}{|c|ccc|}
\hline
$f$ band (Hz) & 40 to 500 & 500 to 1000 & 1000 to 2000 \\
\hline
Tcoh (h) & 60 & 60 & 25 \\
$\delta f$ (Hz) & $3.61\ee{-6}$ & $3.95\ee{-6}$ & $7.75\ee{-6}$ \\
$\delta \dot{f}$ (Hz/s) & $1.16\ee{-10}$ & $1.83\ee{-10}$ & $7.46\ee{-10}$ \\
sky factor & 0.01 & 0.04 & 0.07 \\
$\dot{f}$ refine & 230 & 230 & 150 \\
\hline
min $2\mathcal{F}$ & 6.17 & 6.17 & 5.56 \\
\hline
\end{tabular}
\caption{Einstein@Home MDC search parameters. The sky grid resolution is determined by the sky factor as shown in Equation \ref{eq:EaH_sky}. The spindown resolution used on the fine grid, for the semi-coherent part of the search, is given by $\delta \dot{f}$ divided by the $\dot{f}$-refine value.}
\label{tab:EaH_setup}
\end{table}

\subsection{Frequency Hough}

The Frequency Hough MDC search parameters are given in Table \ref{tab:FH_setup}. The sky grid is constructed in ecliptic coordinates as described in \cite{FH_2}, and is uniform in ecliptic latitude at fixed ecliptic longitude. In a real search the coherent segment length would typically be between 1024 and 8192\,s, depending on the frequency band. For the MDC, it is restricted to 1024\,s over the whole frequency band of the analysis to reduce the computational cost. This implies a sensitivity loss of up to a factor of $\sqrt{8}$ with respect to a real search in the lower frequency bands, where the coherent segment length would typically be higher. 

The Frequency Hough search is performed using all of the MDC data. For the MDC, the search is performed over the whole parameter space. This is because the analysis procedure depends on ranking the most significant candidates per frequency band, as opposed to applying a threshold, so replicating the detection rate of an all-sky search using a smaller parameter space is not trivial. The results for injections where the signal parameters are known are used to optimise the selection of candidates. The search over injections where the signal parameters are unknown are used to validate these results. 

The four most significant candidates per 0.1\,Hz frequency band are selected. A signal is considered detected if there is a candidate within a distance of 3, where the distance is defined in Eq. 10 of \cite{FH_3}. \\

\begin{table}[h!]
\begin{tabular}{|c|c|}
\hline
Tcoh (s) & 1024 \\
$\delta f$ (Hz) & $9.76\ee{-4}$ \\
$\delta \dot{f}$ (Hz/s) & $2.4\ee{-11}$ \\
\hline
\end{tabular}
\caption{Frequency Hough MDC search parameters.}
\label{tab:FH_setup}
\end{table}

\section{Results}
\label{sec:results}

\subsection{Detection efficiency}
\label{sec:res_detection_efficiency}

The detection efficiency, measured on the combined results from the set of known injection and blind injections, is shown in Figure \ref{fig:eff_merged}. (In Appendix \ref{app:stage3_stage4} Figure \ref{fig:eff_stage3_stage4}, we show that the detection efficiency measured using known and blind injections are in agreement.) 

The dependence of the detection efficiency on sensitivity depth is obtained with a sigmoidal fit to the MDC results. The uncertainty band around the resulting sigmoid is obtained by fitting sigmoids to the minimum and maximum of the binomial uncertainties (at the $1 \sigma$ level) on the detection efficiency. The uncertainty band represents the statistical uncertainty on the detection efficiency for this particular search implementation over LIGO S6 data. We would expect to see variations in the measured efficiency with changes in the observation time of the data, quieter or noisier data, search improvements and other changes expected in advanced detector data. When considering the results one should also keep in mind that these are for standard CW signals.

The Frequency Hough results are complete up to 1000\,Hz, however there are no results above this frequency because of technical difficulties with the computer cluster used to perform the search. Specifically, the duration of searches in some frequency bands exceeded the allowance of the cluster for some CPUs. This occurs more at higher frequencies due to the increase in sky grid templates. This issue is being resolved for the Frequency Hough search, but not within the timescale of this MDC. The Frequency Hough results are scaled so that the detection efficiency is measured for the subset of injections for which the search is complete. The results are displayed with a hatched uncertainty band to highlight the difference with respect to the other searches. 
    
Figure \ref{fig:eff_merged} shows that the results from the Frequency Hough, Sky Hough and Powerflux searches are comparable. If we compare the sensitivity depth achieved at 60\% efficiency we see that Time domain $\mathcal{F}$-statistic is less sensitive to these standard CW signals. At the same efficiency, the Einstein@Home search is a factor of two more sensitive than the next most sensitive search. This difference can be attributed to a combination of the significant computing resources of the Einstein@Home project, the longer coherent segment length, recent method improvements, and the intensive refinement procedure which allows for the follow-up of many candidates from the all-sky search. 

In Figure \ref{fig:eff_merged} it is clear that the detection efficiency does not reach 100\% for very strong signals. For Einstein@Home, this is due to signals which overlap with known noise lines, as shown in Section \ref{sec:results_noise}. For Powerflux, this due to signals below 400\,Hz, as shown in \ref{sec:results_frequency}.\\

\begin{figure}[htb!]
  \includegraphics[width=3.5in]{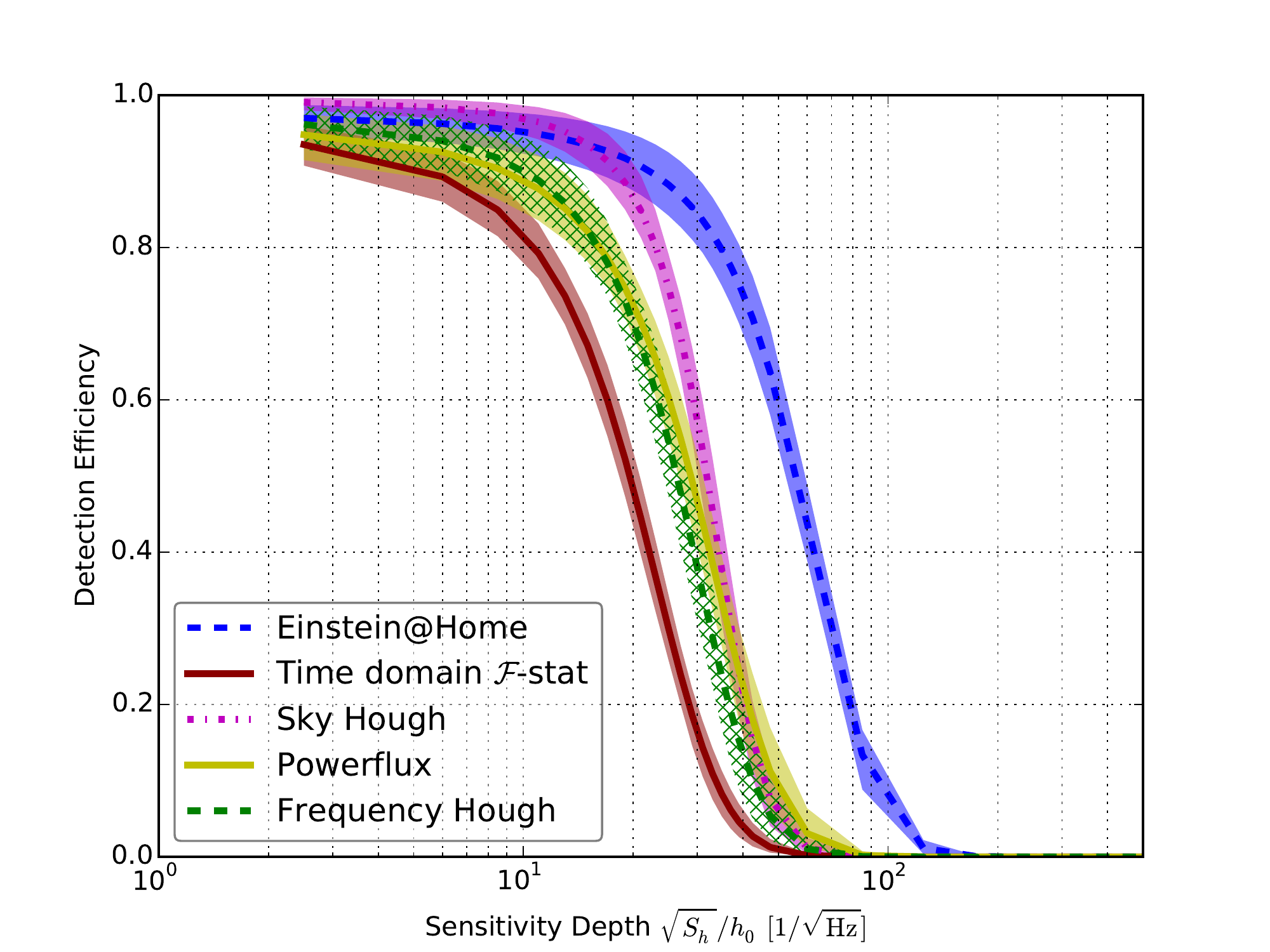}\\
\caption{\label{fig:eff_merged} Detection efficiency measured for all 3110 injections. The Frequency Hough results are shown with a hatched error band because the efficiency is measured for a subset (1920) of the MDC injections. The curves and error bands are obtained by fitting sigmoids to the data, see Section \ref{sec:res_detection_efficiency}. The error bands represent the statistical uncertainty on the detection efficiency measured for the search implemention and data used in the MDC.}
\end{figure}   

\subsubsection{Robustness in the presence of detector artifacts}
\label{sec:results_noise}

\begin{figure}[htb!]
  \includegraphics[width=3.5in]{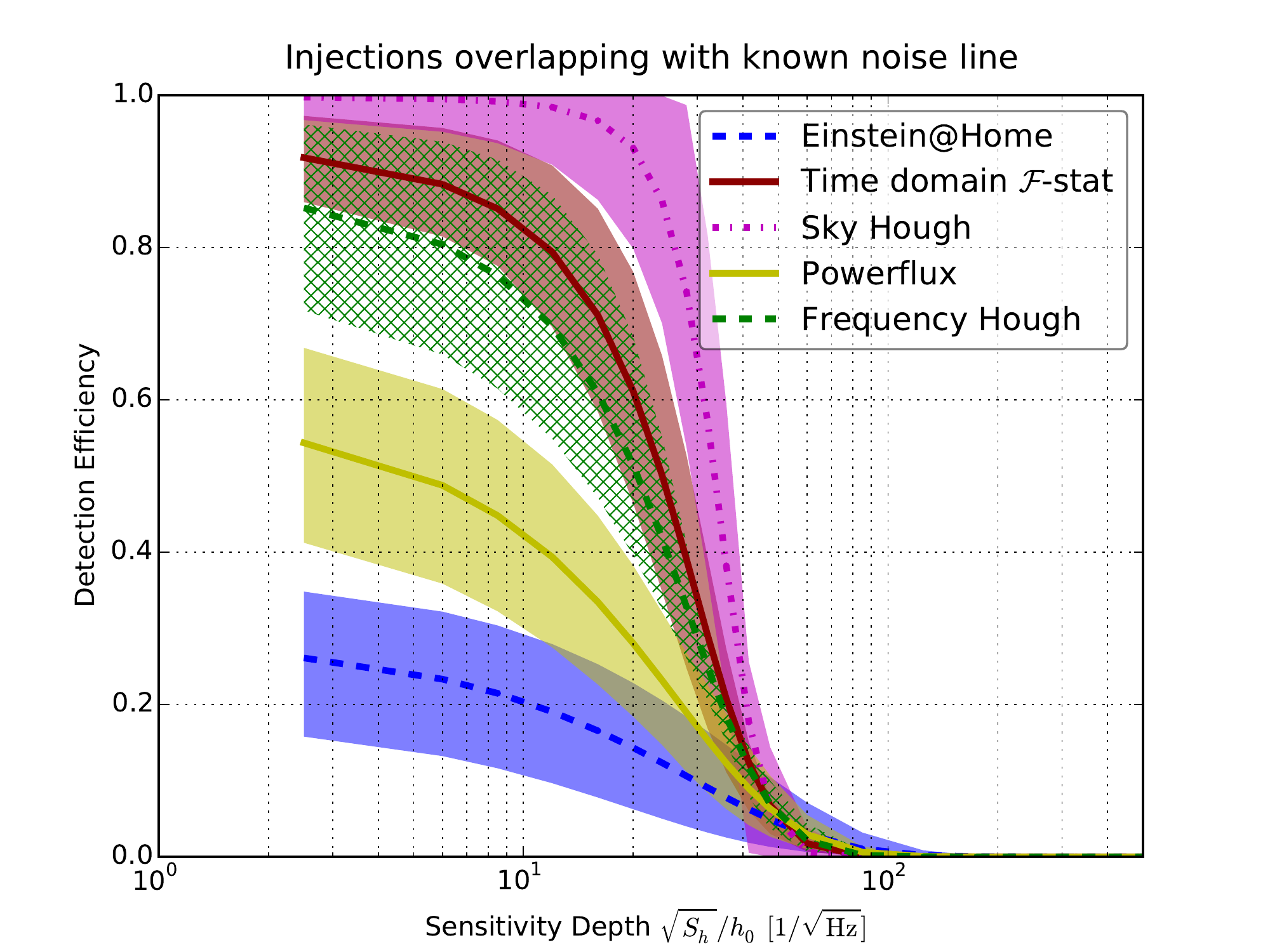}\\
  \includegraphics[width=3.5in]{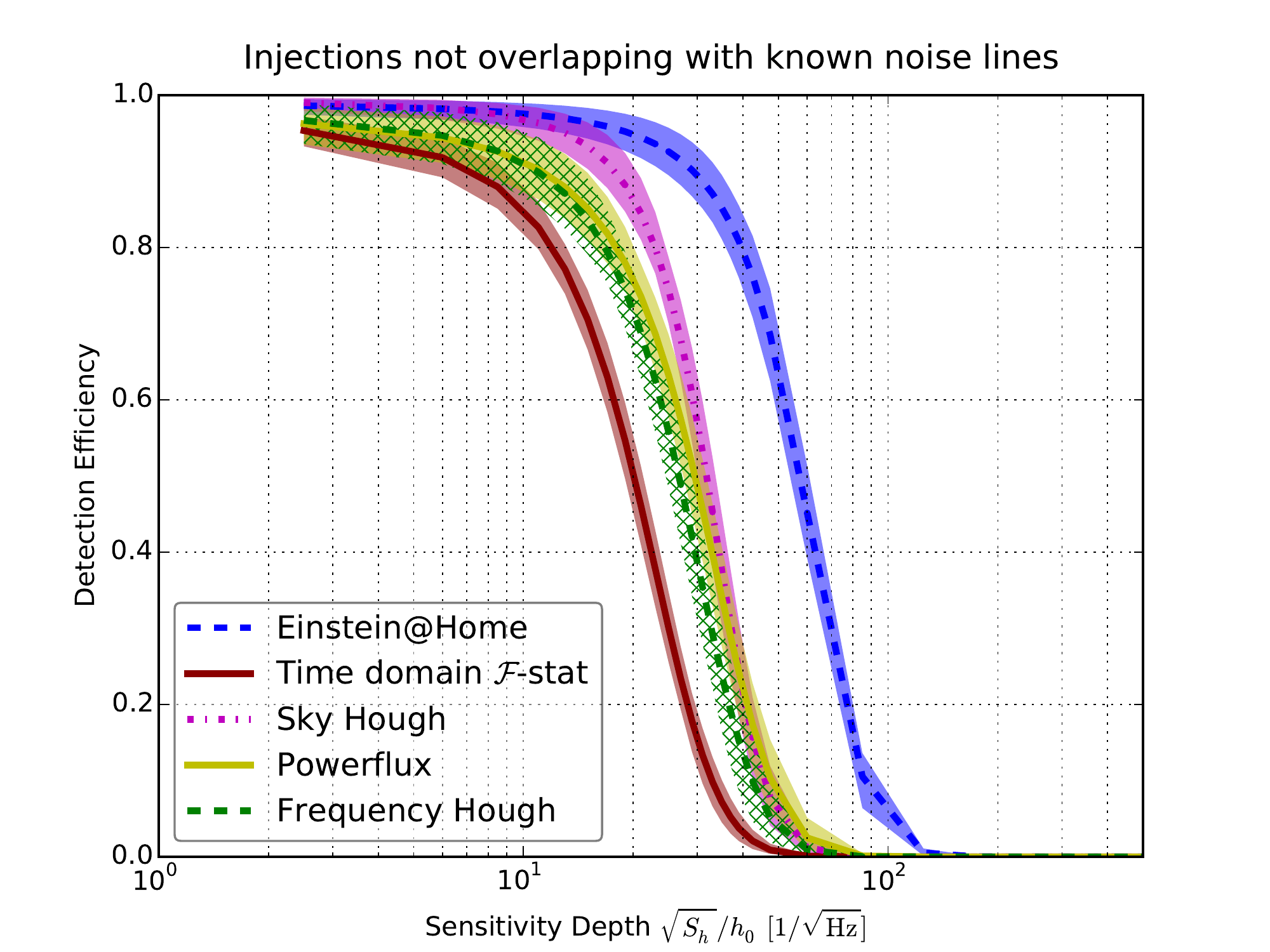}\\
\caption{\label{fig:eff_merged_nkl} Detection efficiency measured for injections overlapping with known noise lines (top, 184 injections), and when injections which overlap with known noise lines are excluded (bottom, 2926 injections). The Frequency Hough results are shown with a hatched error band because the efficiency is measured for a subset of the MDC injections (top: 117, bottom: 1803). The curves and error bands are obtained by fitting sigmoids to the data, see Section \ref{sec:res_detection_efficiency}. The error bands represent the statistical uncertainty on the detection efficiency measured for the search implemention and data used in the MDC.}
\end{figure}

Each method has a different procedure for excluding candidates caused by detector artifacts, also known as noise lines, described in Section \ref{sec:methods}. In Figure \ref{fig:eff_merged_nkl} we separate the detection efficiency measured in quiet data, and the detection efficiency measured for injections whose frequency overlaps with known noise lines.

The top panel of Figure \ref{fig:eff_merged_nkl} shows that the efficiency for the Sky Hough, Time domain $\mathcal{F}$-statistic and Frequency Hough searches remains unchanged, within the measurement uncertainty, in the presence of noise. The Frequency Hough procedure for handling lines are not included in the MDC, therefore these results are not representative of noise handling in a real search. 

As Einstein@Home applies an aggressive cleaning procedure, where known noise lines are replaced by Gaussian noise, any signal which overlaps with a noise line in both detectors is removed. When signal overlaps with a noise line in one detector, the signal is suppressed by the logBSGL statistic which downweights signal appearing in one detector. In the case of Powerflux, signals overlapping with noise lines is suppressed by the procedure where SFTs are weighted according to their noise level. 

The bottom panel of Figure \ref{fig:eff_merged_nkl} shows that, in the absence of known lines, the efficiency for strong signals has increased for Powerflux and reaches almost 100\% for Einstein@Home. \\
 
\subsubsection{Dependence on signal frequency or spindown}
\label{sec:results_frequency}

\begin{figure}[htb!]
  \includegraphics[width=3.2in]{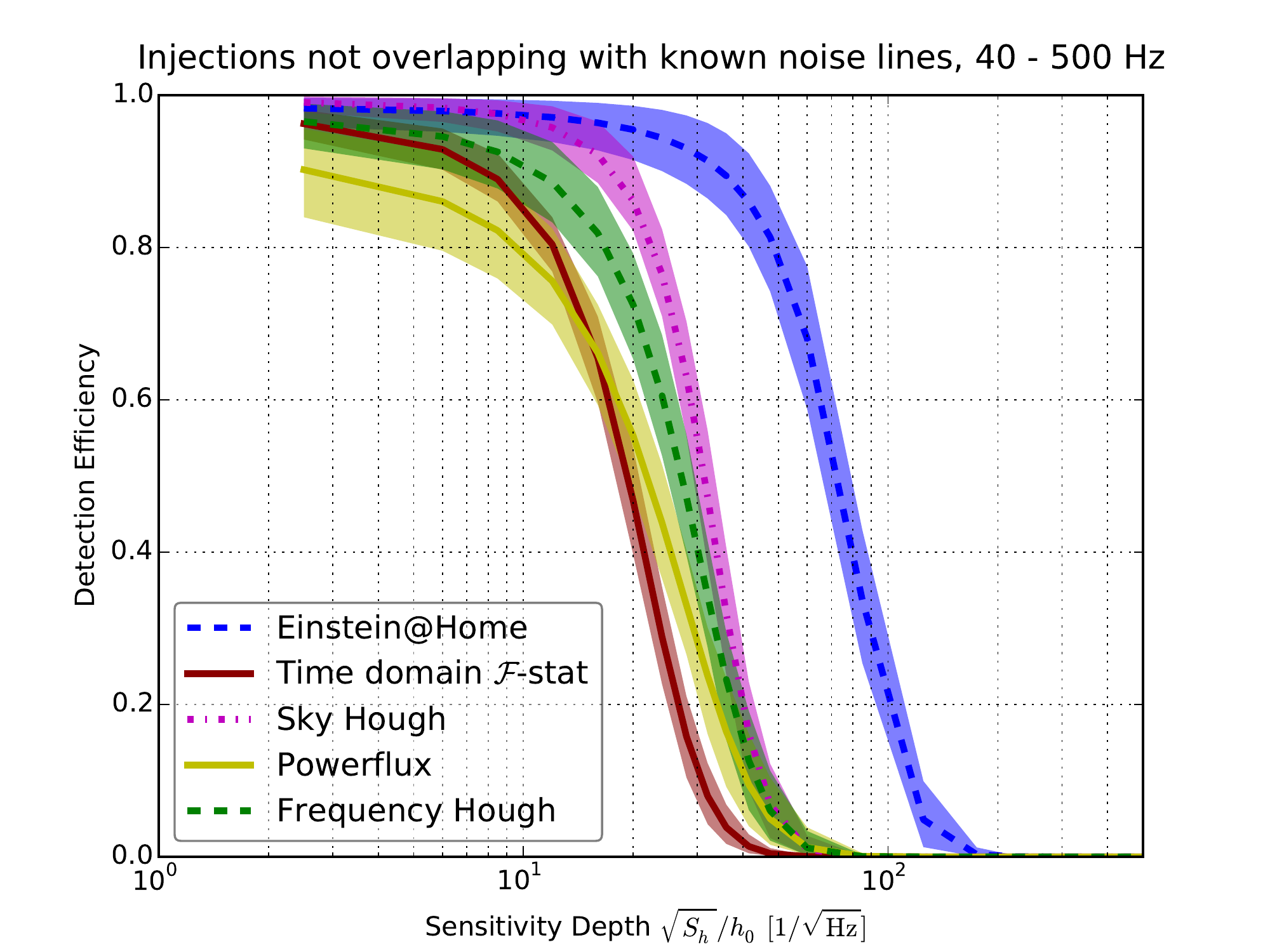}\\
  \includegraphics[width=3.2in]{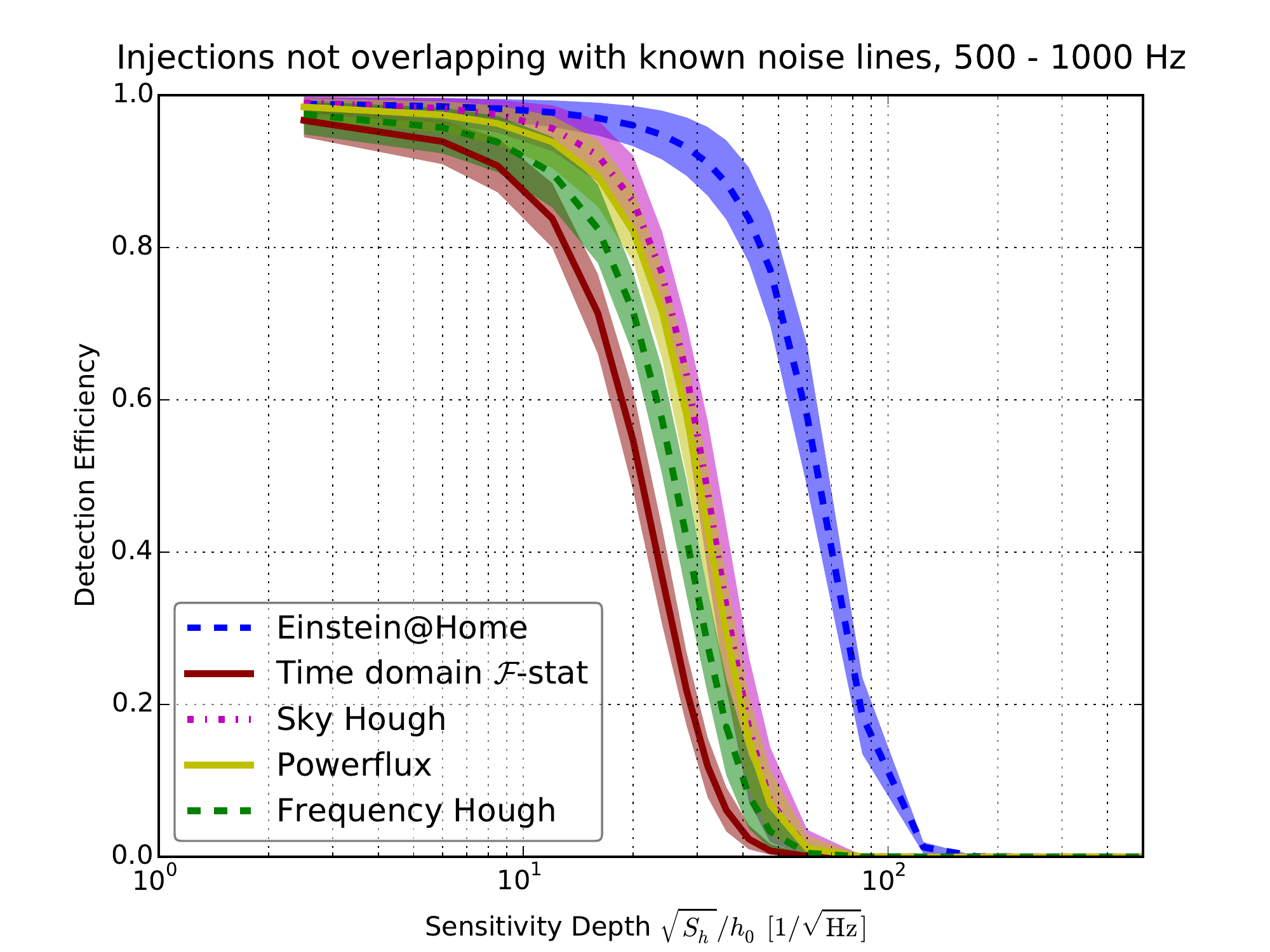}\\
  \includegraphics[width=3.2in]{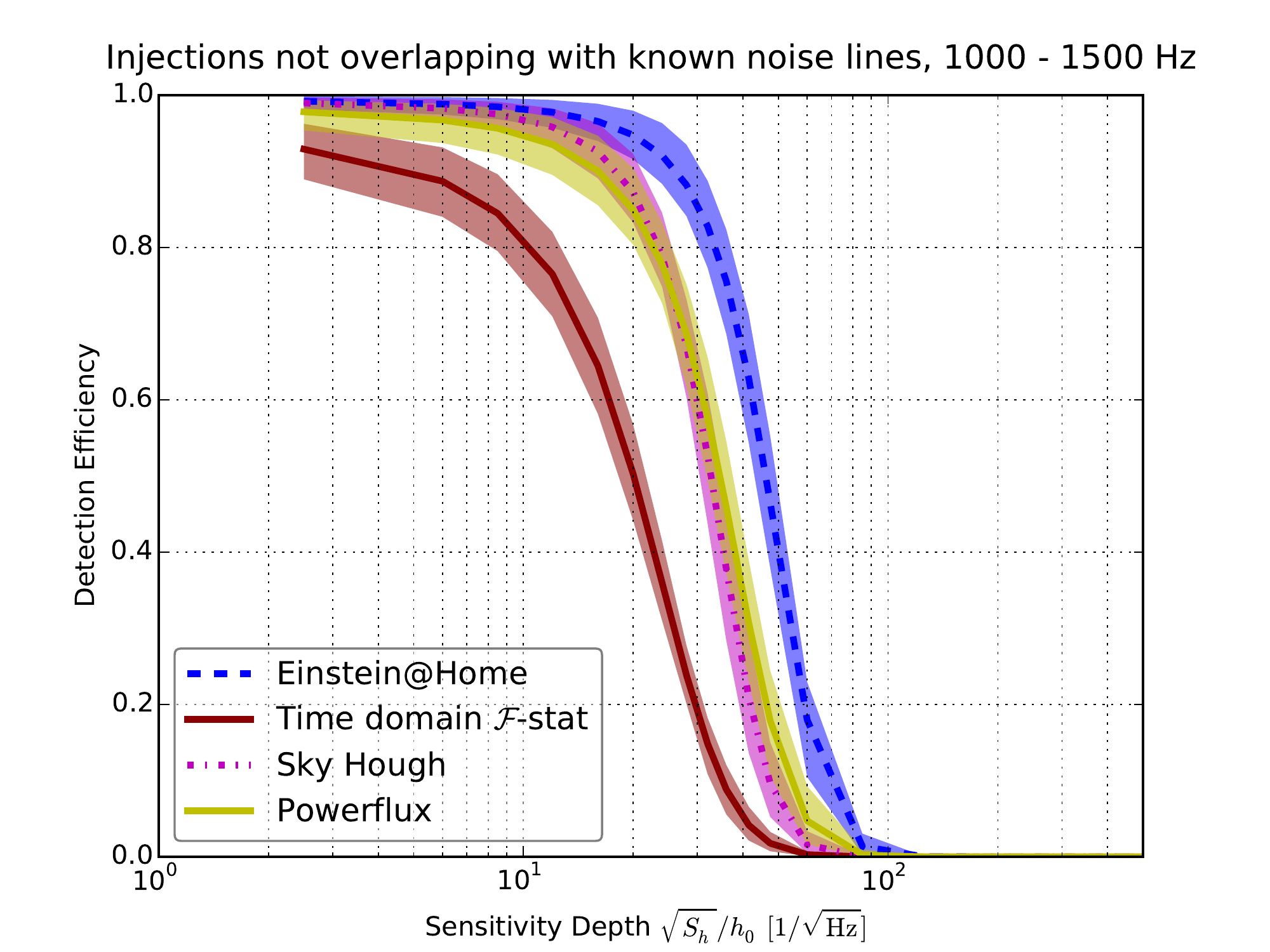}\\
\caption{\label{fig:eff_merged_splitf0} Detection efficiency measured for injections in the the frequency ranges of 40 to 500\,Hz, 500 to 1000\,Hz and 1000 to 1500\,Hz (859, 944, and 1123 injections respectively). The Frequency Hough results are complete for injections below 1000\,Hz. The curves and error bands are obtained by fitting sigmoids to the data, see Section \ref{sec:res_detection_efficiency}. The error bands represent the statistical uncertainty on the detection efficiency measured for the search implemention and data used in the MDC.}
\end{figure}

Here we consider the detection efficiency only for injections that do not overlap with known noise lines. Figure \ref{fig:eff_merged_splitf0} shows the detection efficiency separately for injections in the frequency ranges of 40 to 500\,Hz, 500 to 1000\,Hz and 1000 to 1500\,Hz.

The Sky Hough, Time domain $\mathcal{F}$-statistic and Frequency Hough results do not depend on frequency. This indicates the Frequency Hough results would not change if injections above 1000\,Hz were included. Powerflux measures lower efficiency in the low frequency range. This is expected as the S6 analysis applied in the MDC was only tuned for signals above 400\,Hz. In the higher frequency bands, for which the search is designed, the detection efficiency approaches 100\% for the strongest signals.
 
The drop in efficiency for the Einstein@Home search at higher frequencies is expected due to the choice of having equal computing cost assigned to the searches in each of the three frequency bands in Table \ref{tab:EaH_setup}. As the frequency increases a higher sky-grid density, and therefore a higher computing cost, is required to achieve the same sensivity. In order to keep the computing cost fixed, a coarser search grid is used in the higher frequency bands.\\

\begin{figure}[htb!]
  \includegraphics[width=3.2in]{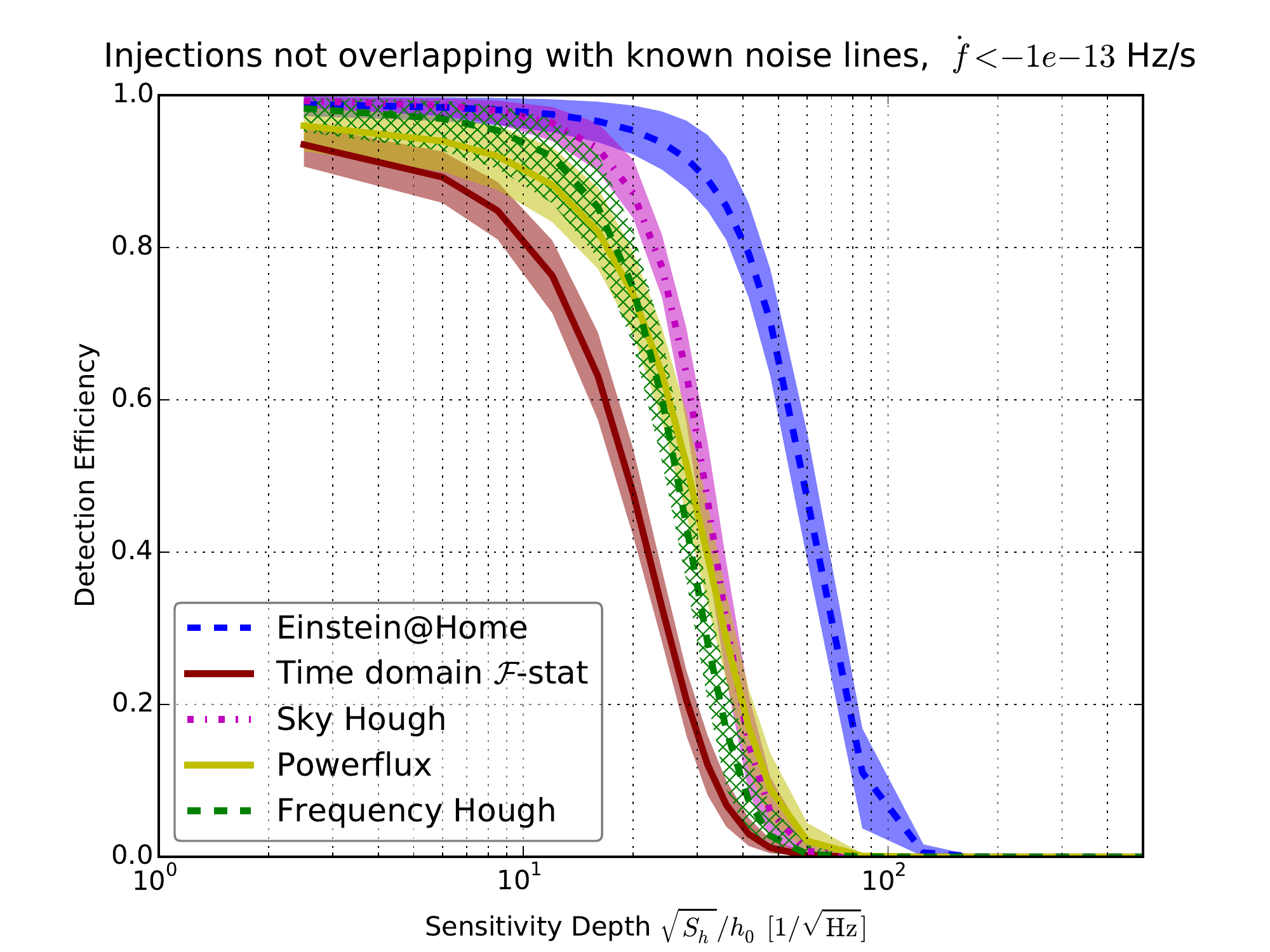}\\
  \includegraphics[width=3.2in]{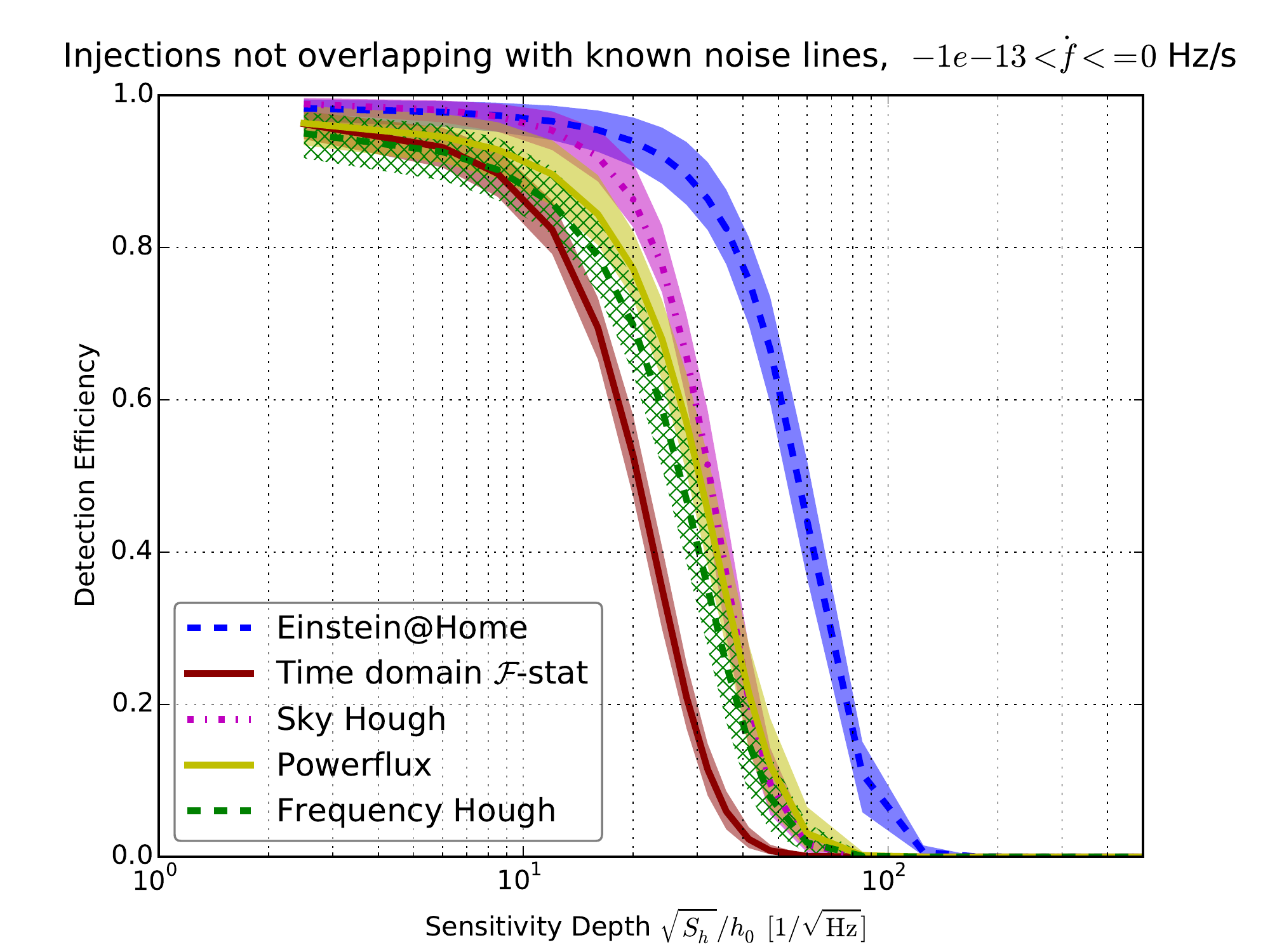}\\
  \includegraphics[width=3.2in]{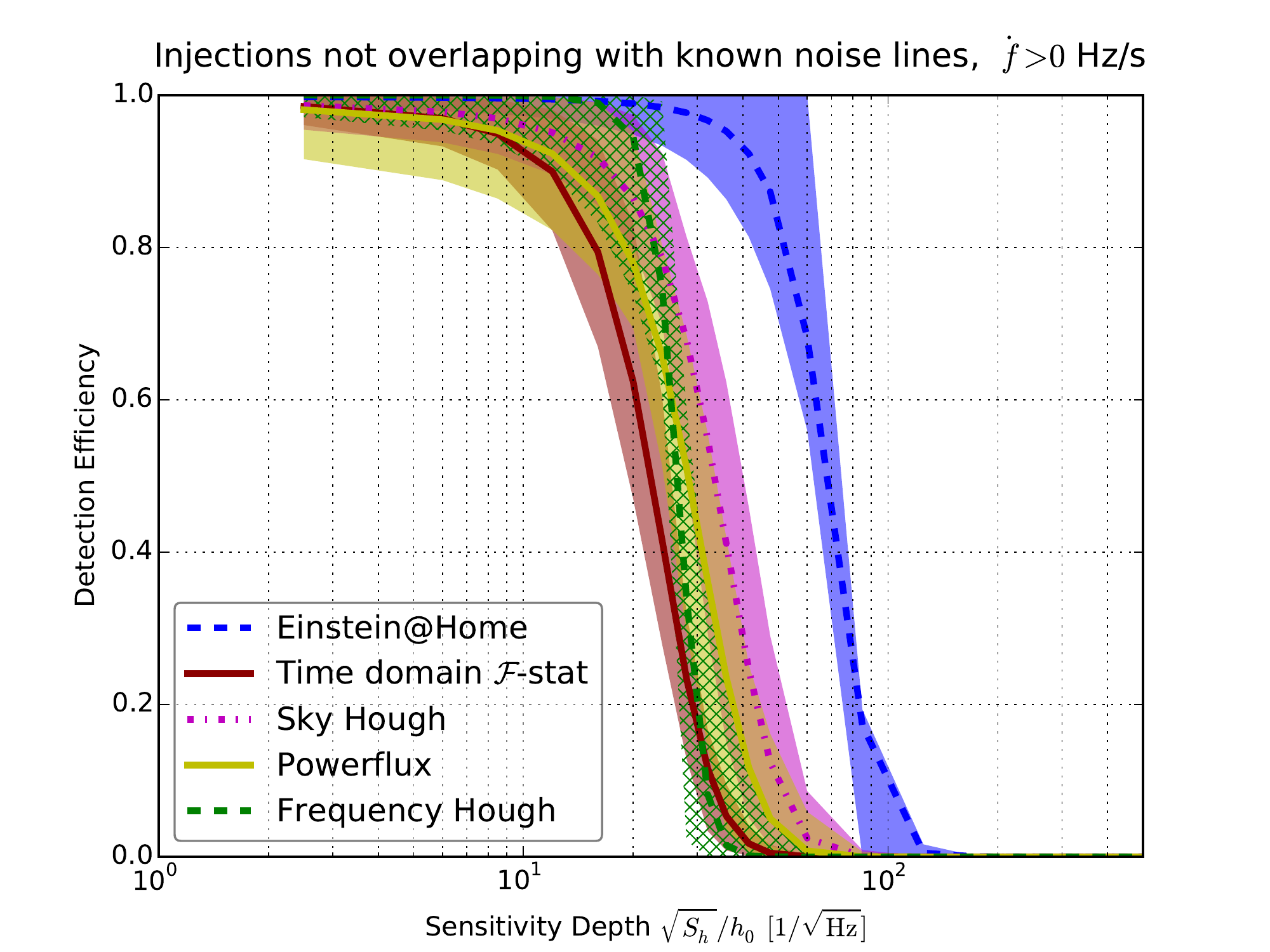}\\
\caption{\label{fig:eff_merged_splitf1} Detection efficiency measured for injections with small spindown ($< -1\ee{-13}$\,Hz/s, 1260 injections), large spindown ($-1\ee{-13}$ to 0\,Hz/s, 1517 injections) and with spinup ($>$ 0\,Hz/s, 149 injections). Frequency Hough results are shown with a hatched error band because the efficiency is measured for a subset of the MDC injections (793, 919, and 457 injections respectively). The curves and error bands are obtained by fitting sigmoids to the data, see Section \ref{sec:res_detection_efficiency}. The error bands represent the statistical uncertainty on the detection efficiency measured for the search implemention and data used in the MDC.}
\end{figure}

Figure \ref{fig:eff_merged_splitf1} shows the detection efficiency for injections with large spindown, small spindown and with spinup. There is no dependence on the frequency derivative of the signal for any of the searches. \\

\subsubsection{Dependence on signal second order spindown}
\label{sec:results_second_order_spindown}

A loss in detection efficiency is expected when the signal $\ddot{f}$ is greater than the $\ddot{f}_\mathrm{critical}$ for a search, where $\ddot{f}_\mathrm{critical}$ is given by

\begin{equation}
\ddot{f}_\mathrm{critical} = \frac{\delta{f}}{\mathrm{T_{obs}}^2}.
\end{equation}
$\ddot{f}_\mathrm{critical}$ is the value of $\ddot{f}$ at which the signal frequency will vary by more than a frequency bin, $\delta{f}$, over the observation time of the data, $\mathrm{T_{obs}}$. In practice, the efficiency loss for $\ddot{f} \ge \ddot{f}_\mathrm{critical}$ is expected to be mitigated to some degree by apparent displacement of the signal parameters in the space of $(f_0,\dot{f},\alpha,\delta)$.

\begin{figure}[htb!]
  \includegraphics[width=3.5in]{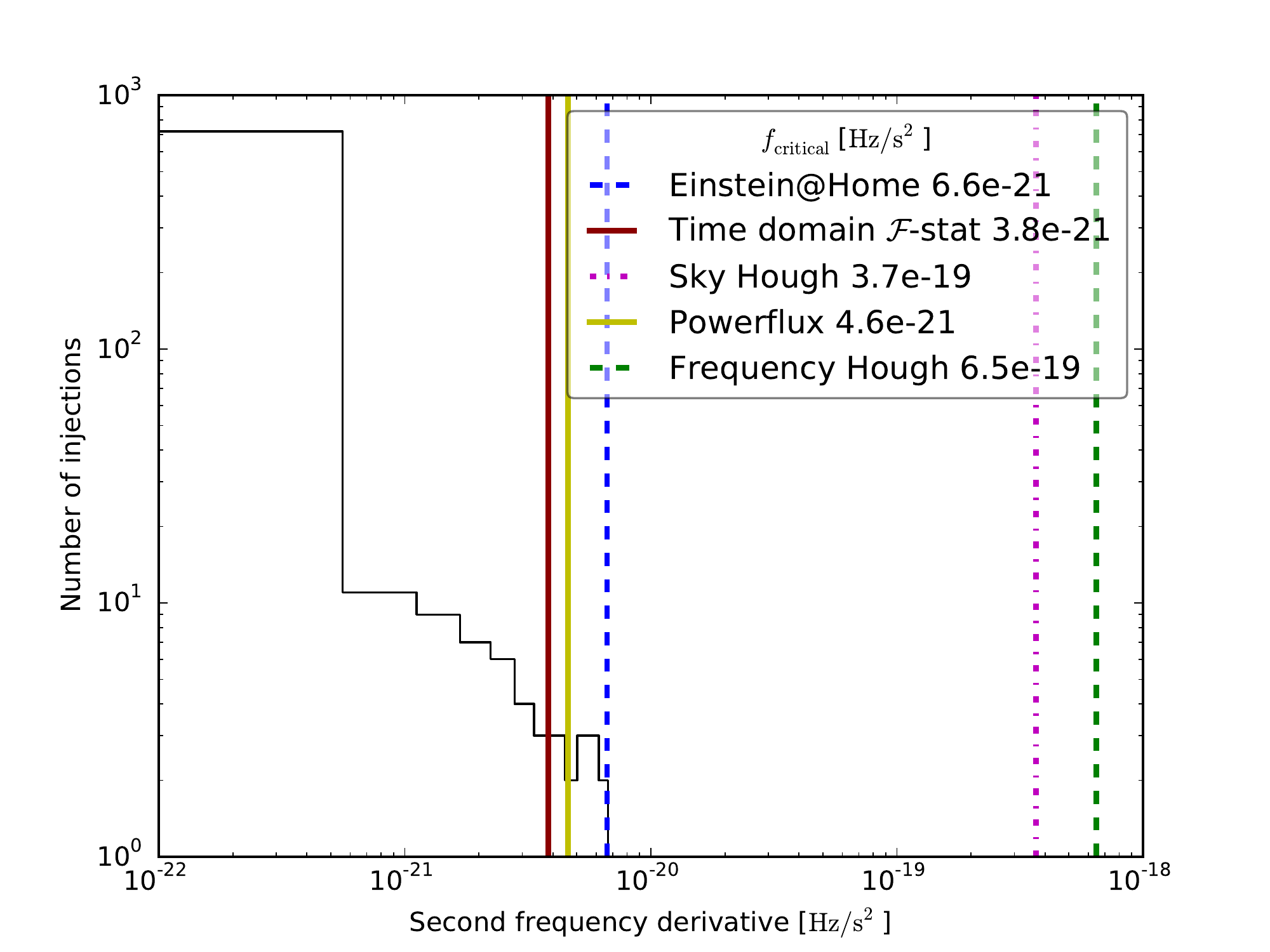}
\caption{\label{fig:f2_injections_fcrit} Distribution of $\ddot{f}$ values for 781 injections with $\ddot{f} > 0$. The vertical lines show $\ddot{f}_\mathrm{critical}$ for each of the pipelines. The $\ddot{f}_\mathrm{critical}$ is calculated after the first refinement stage for Powerflux and for the 40 to 500\,Hz search setup for Einstein@Home.}
\end{figure}

Figure \ref{fig:f2_injections_fcrit} shows the non-zero second order spindown values of the MDC signals, for the range specified in Section \ref{sec:the_data}. The vertical lines show the $\ddot{f}_\mathrm{critical}$ for each of the searches. Some signals have $\ddot{f} \ge \ddot{f}_\mathrm{critical}$ for the Einstein@Home, Time domain $\mathcal{F}$-statistic and Powerflux searches. However, in each case, there are too few signals with $\ddot{f} \ge \ddot{f}_\mathrm{critical}$ to determine if they have an appreciable effect on the detection efficiency.

\begin{figure}[htb!]
  \includegraphics[width=3.5in]{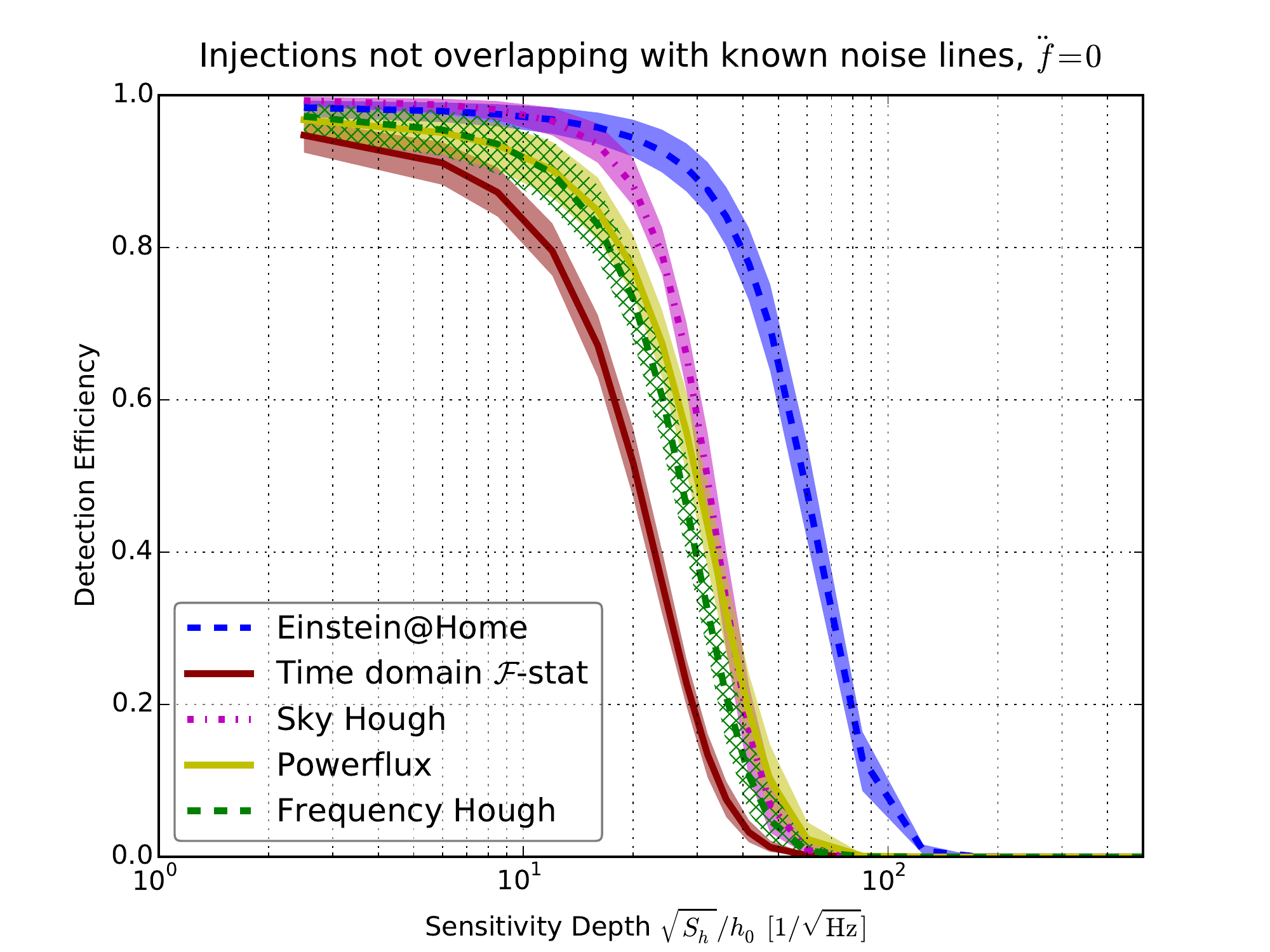}\\
  \includegraphics[width=3.5in]{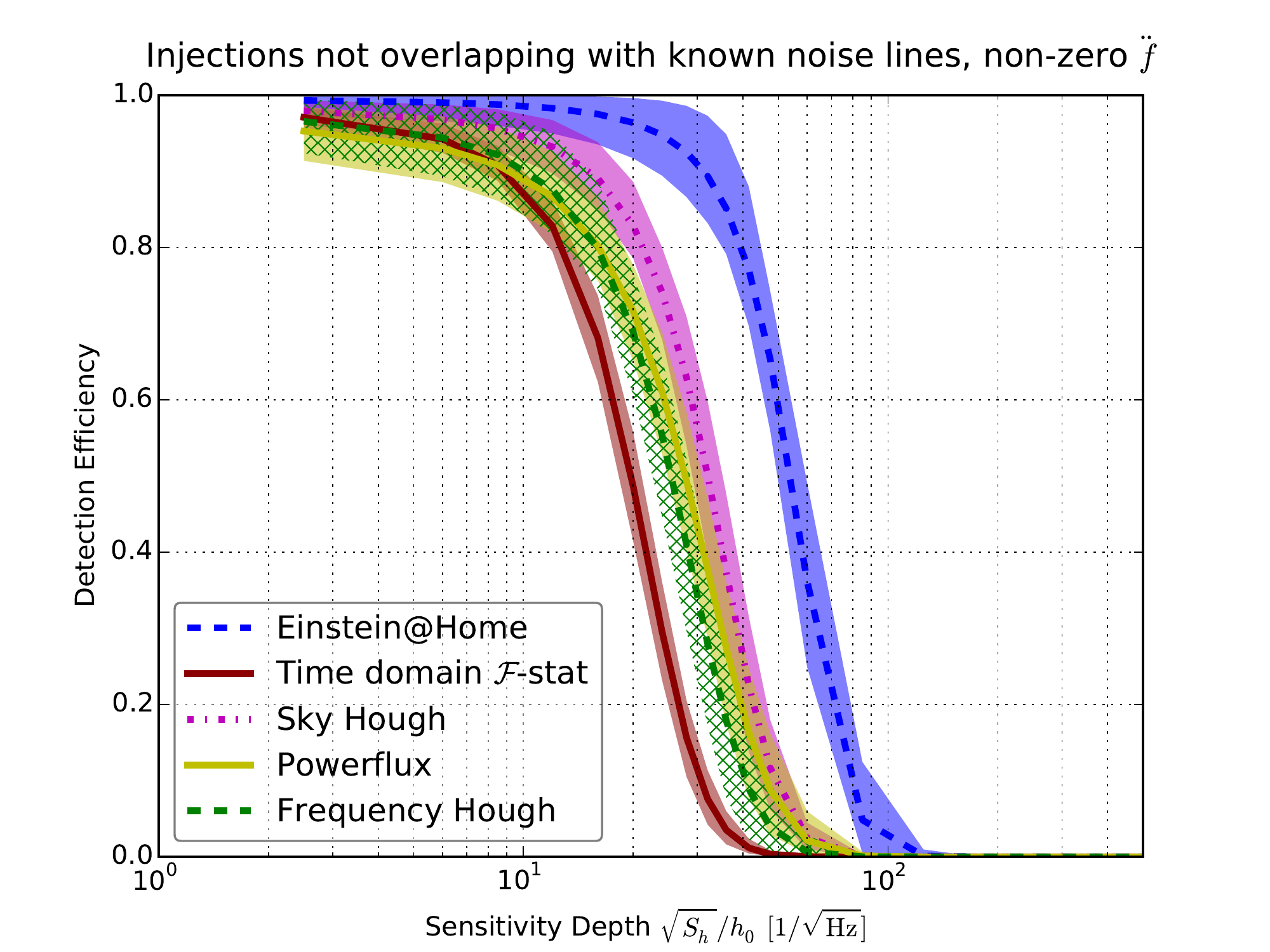}\\
\caption{\label{fig:eff_merged_splitf2} Detection efficiency measured for injections with zero (2329 injections) and non-zero (781 injections) second order spindown. Frequency Hough results are shown with a hatched error band because the efficiency is measured for a subset of the MDC injections. The curves and error bands are obtained by fitting sigmoids to the data, see Section \ref{sec:res_detection_efficiency}. The error bands represent the statistical uncertainty on the detection efficiency measured for the search implemention and data used in the MDC.}
\end{figure}

The stability of the detection efficiency for signals with $\ddot{f} > 0$ is important as none of the pipelines search explicity over second order spindown, and to do so would add a significant computational burden to the searches. Figure \ref{fig:eff_merged_splitf2} shows that the detection efficiency is the same for signals with $\ddot{f} = 0$ and $\ddot{f} > 0$, with at least $99\%$ of injections having $\ddot{f} < \ddot{f}_\mathrm{critical}$. Due to the lack of injections with $\ddot{f} \ge \ddot{f}_\mathrm{critical}$, the impact on detection from these injections not examined. 

For the Sky Hough and Frequency Hough searches the impact of $\ddot{f} \ge \ddot{f}_\mathrm{critical}$ is less of a concern for future searches, due to the short coherent segment length used by these searches. The $\ddot{f}_\mathrm{critical}$ of the Time domain $\mathcal{F}$-statistic search will be larger than the value in the MDC for the first advanced LIGO searches, which will have a lower $\mathrm{T_{obs}}$ than the 15 months of the MDC data.

The Powerflux $\ddot{f}_\mathrm{critical}$ is calculated for refinement stage 1 in Table \ref{tab:PF_setup}. The $\delta{f}$ decreases in the next refinement stages, so $\ddot{f}_\mathrm{critical}$ will decrease. The Einstein@Home search refinement stages also have reduced $\delta{f}$ and $\ddot{f}_\mathrm{critical}$. Therefore, the impact on detection efficiency of $\ddot{f} \ge \ddot{f}_\mathrm{critical}$ may warrant further study for these two searches.\\

\subsection{Signal parameter recovery}

The distance between the signal and recovered candidates is of interest as it determines the region in parameter space around each candidate that must be searched by the next refinement stage. Also, it serves as a useful cross check that candidates being claimed as detections in the MDC are within a reasonable distance of the signal parameters. 

Powerflux, Einstein@Home and Frequency Hough can have more than one candidate per injection. For strong signals, there are in fact many detection candidates around the signal's true parameter values. We study the distribution for the distances of candidates from the true signal parameter values for two sets of candidates: the one with the highest SNR and the one that is closest, in frequency, to the signal.

\begin{figure}[htb!]
  \includegraphics[width=3.2in]{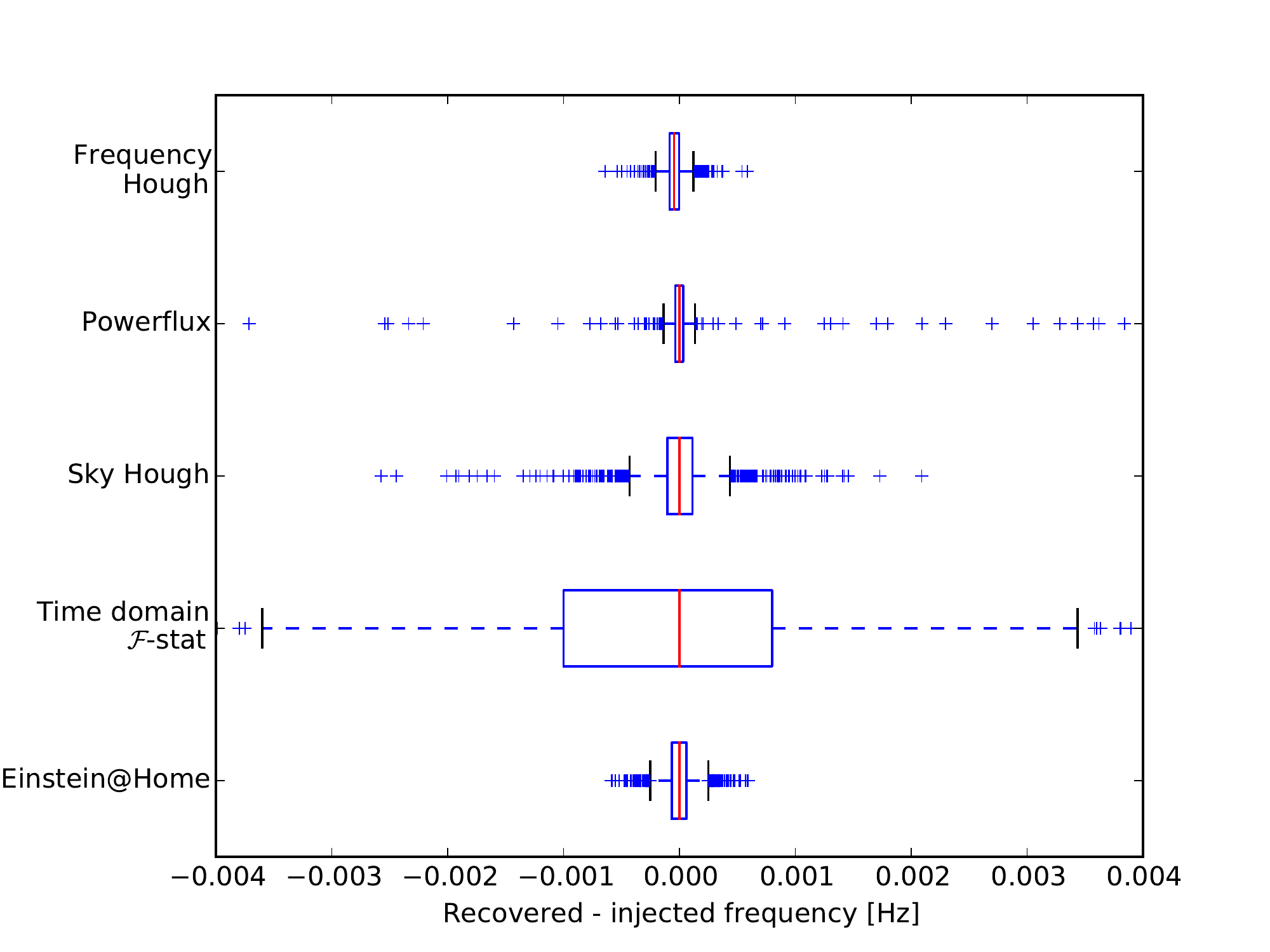}\\
  \includegraphics[width=3.2in]{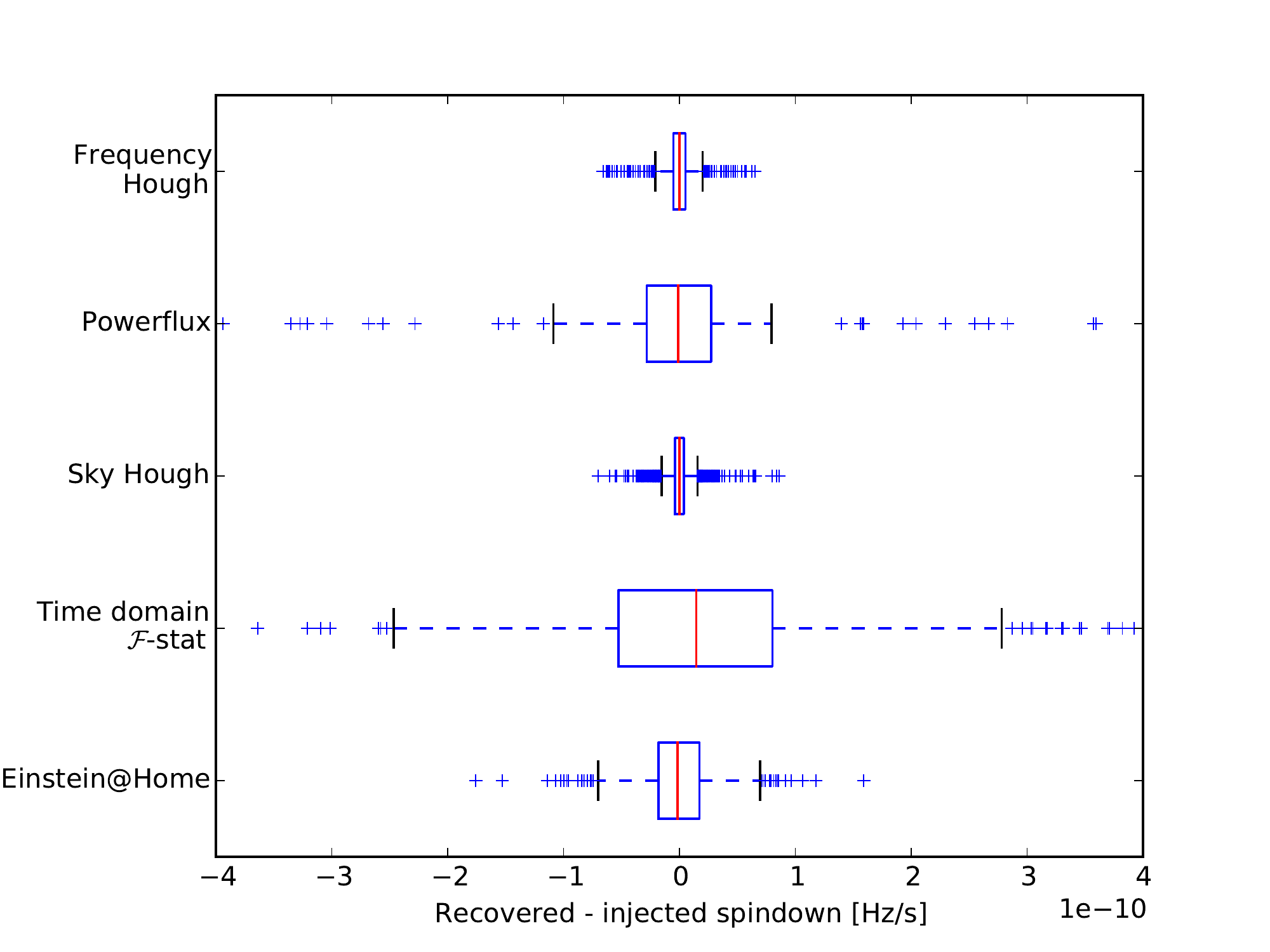}\\
  \includegraphics[width=3.2in]{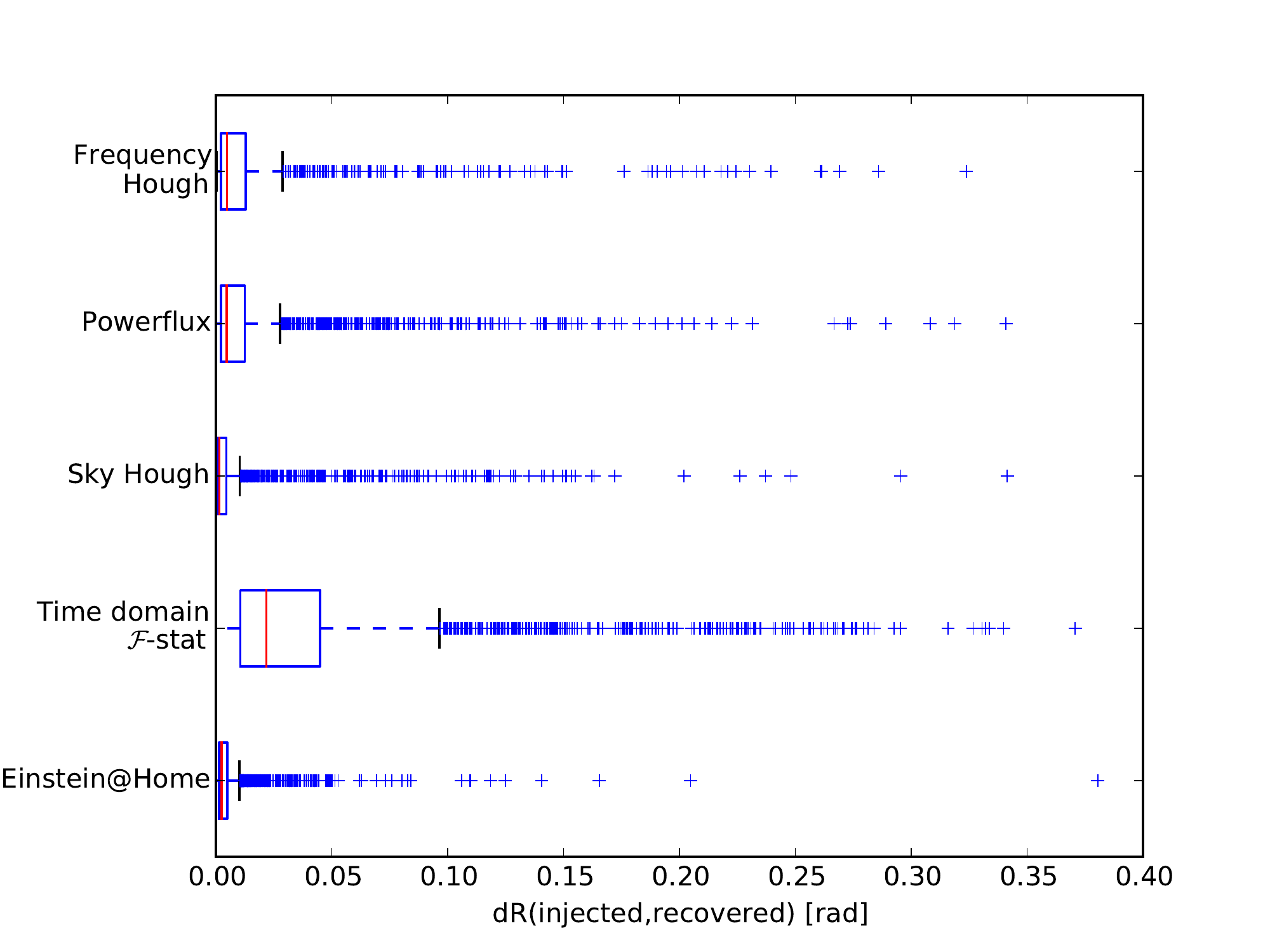}\\

\caption{\label{fig:distances_merged} The distance between the signal and the recovered candidate, in frequency, spindown and sky position, when the candidate with the highest SNR is chosen. The red line is the median. The blue box begins and ends at the first and third quartile respectively. The vertical black bars (whiskers) extend 1.5 times the inner quartile range outside the blue box. The blue crosses are candidates outside this range.}
\end{figure}

\begin{figure}[htb!]
  \includegraphics[width=3.2in]{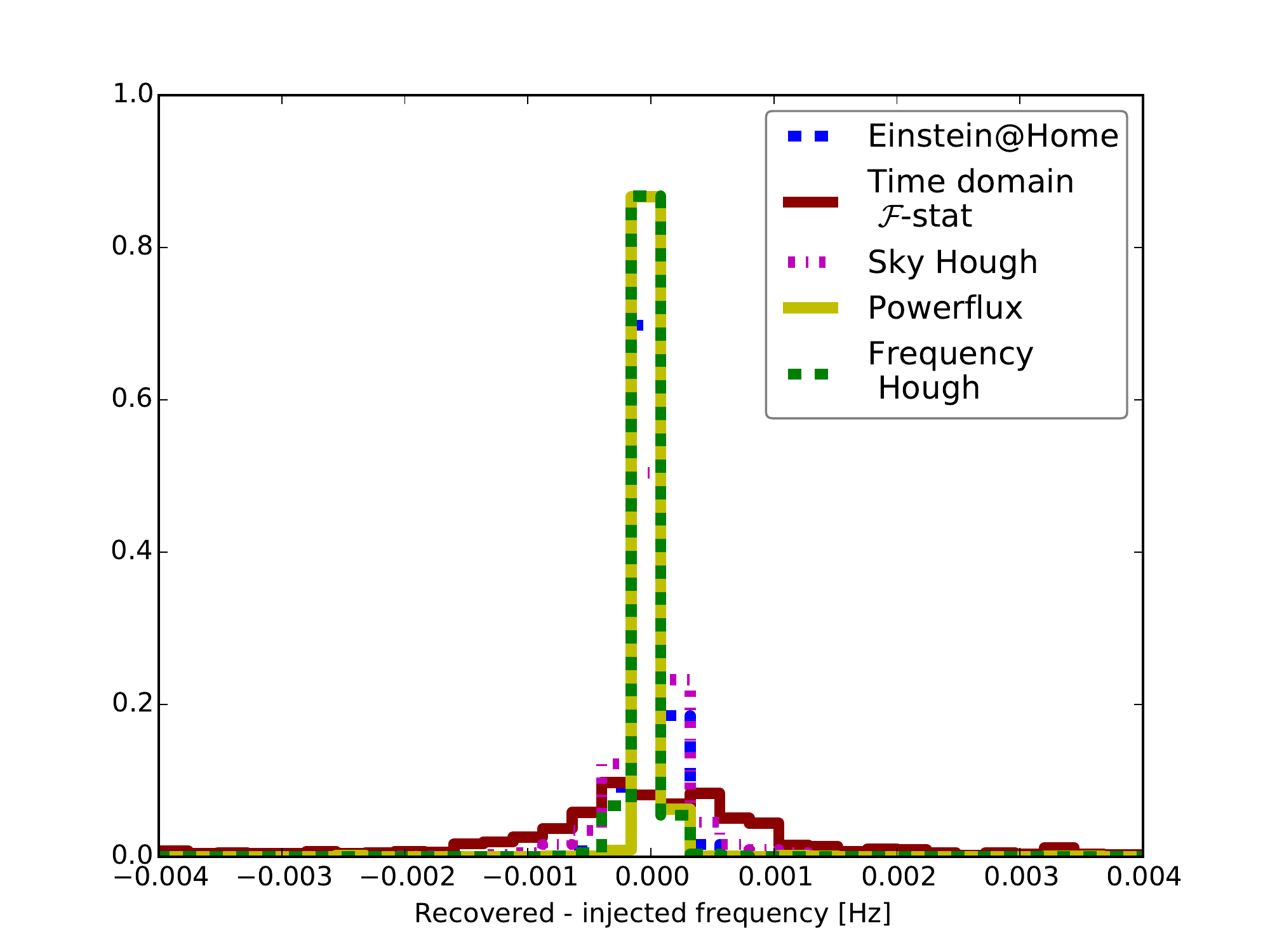}\\
  \includegraphics[width=3.2in]{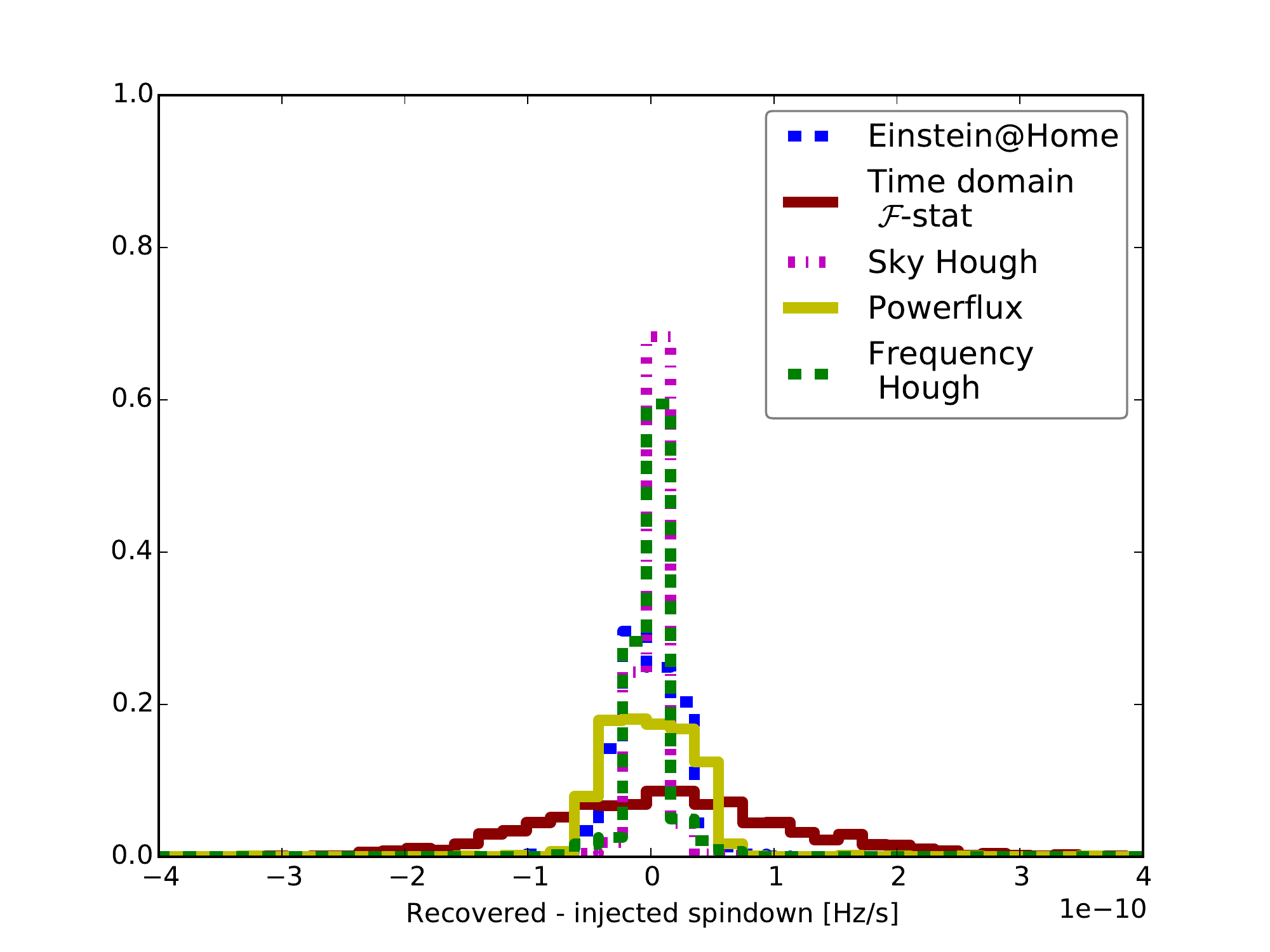}\\
  \includegraphics[width=3.2in]{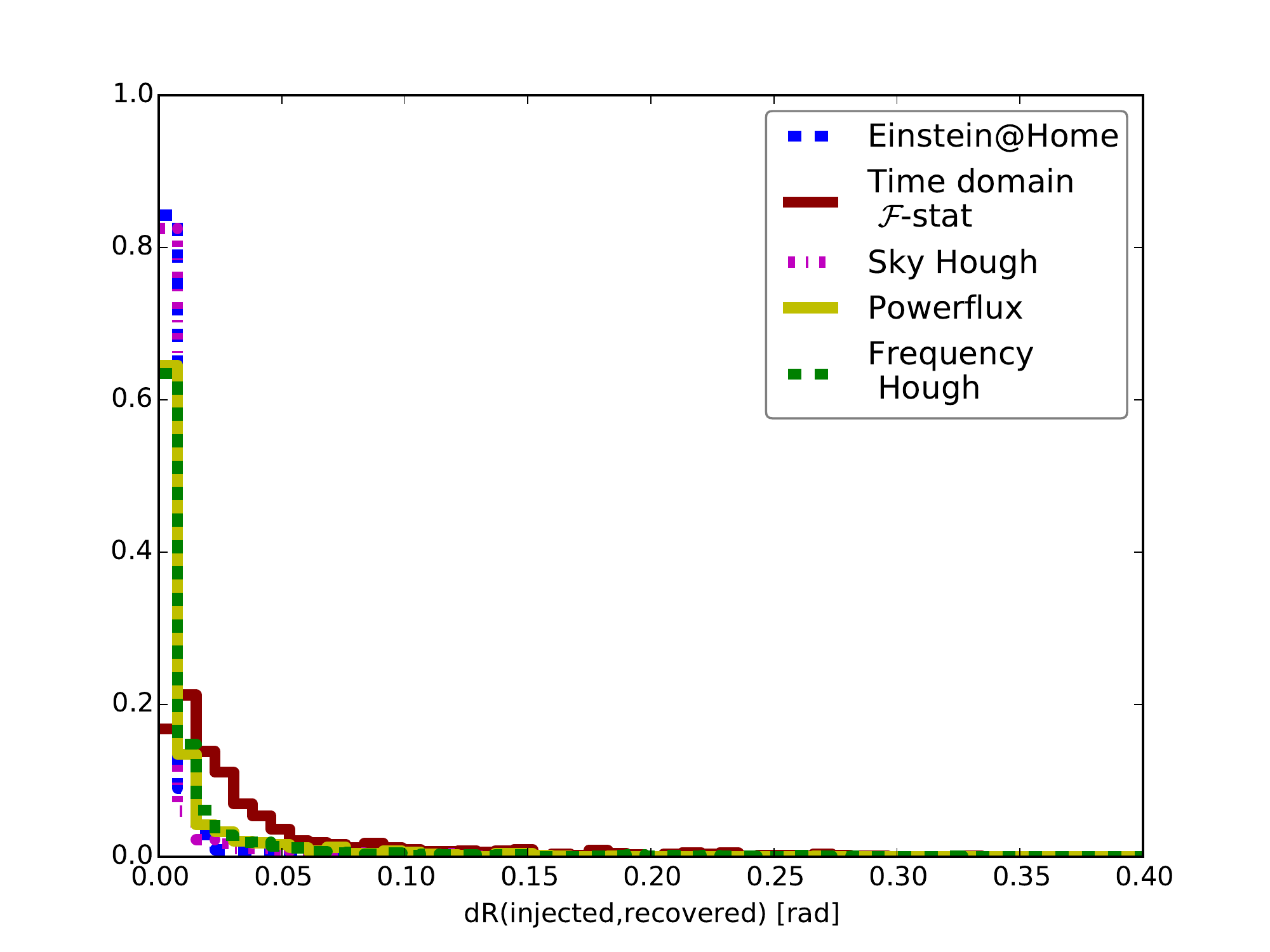}\\

\caption{\label{fig:distances_merged_hist} The distance between the signal and the recovered candidate, in frequency, spindown and sky position, when the candidate with the highest SNR is chosen. This is an alternative representation of the same data as in Figure \ref{fig:distances_merged}.}
\end{figure}

The distance, in frequency, spindown and sky position, is shown for the first set of candidates in Figures \ref{fig:distances_merged} and \ref{fig:distances_merged_hist}. Here the distance in sky is represented by dR, the angular separation between two sky positions in radians. The dR scales approximately proportional to the frequency of the signal. There are candidates from Powerflux and Time domain $\mathcal{F}$-statistic which are not shown because they lie outside the limits of the x-axis. For Powerflux, these amount to 4\%, 5\% and 0.6\% of candidates outside the boundaries in frequency, spindown and dR, with $< 0.1\%$ of candidates outside all three boundaries. For Time domain $\mathcal{F}$-statistic, 27\% are $> 4\ee{-3}$\,Hz from the signal while 3\% are $> 4\ee{-10}$\,Hz/s, no candidates are outside both boundaries. The Powerflux method recovers candidates up to 0.02\,Hz, $1.5\ee{-9}$\,Hz/s or dR $= 1.5$ from the injected parameters. The Time domain $\mathcal{F}$-statistic recovers candidates up to 0.06\,Hz, $1.5\ee{-9}$\,Hz/s or dR $= 0.4$ from the injection. This means the region these searches need to search to recover the signal from these candidates is larger by a factor of O(10) in each dimension than the other searches. Time domain $\mathcal{F}$-statistic can afford to do this because they expect $\sim 10$ false alarms. Powerflux expects on the order of 10000 false alarms in an all-sky search. While Powerflux may refine more than one candidate from a signal, only one of the candidates needs to pass through the refinement stages in order to recover the signal. Therefore, the minimum search region needed to recover the signal is better representated by examining the closest signal in frequency. 

The Einstein@Home and Frequency Hough refinement searches, on candidates from the initial search, cover a predefined region in parameter space around each candidate. In a real search, candidates from a signal within this region will result in a detection after refinement. Therefore, the MDC detection candidates are required to contain the signal within this region (Section \ref{sec:implementation}). Figures \ref{fig:distances_merged} and \ref{fig:distances_merged_hist} support these choices of refinement parameter space, as the majority of candidates are within a smaller region around the signal and there are few outliers at the bounds of the parameter space. 

The Sky Hough search expects $< 15000$ false alarms in an all-sky search over 1500\,Hz, because at most one candidate per 0.1\,Hz is selected. The recovered parameters are close to the signal parameters. This allows for quick turnover of results. 

\begin{figure}[htb!]
  \includegraphics[width=3.2in]{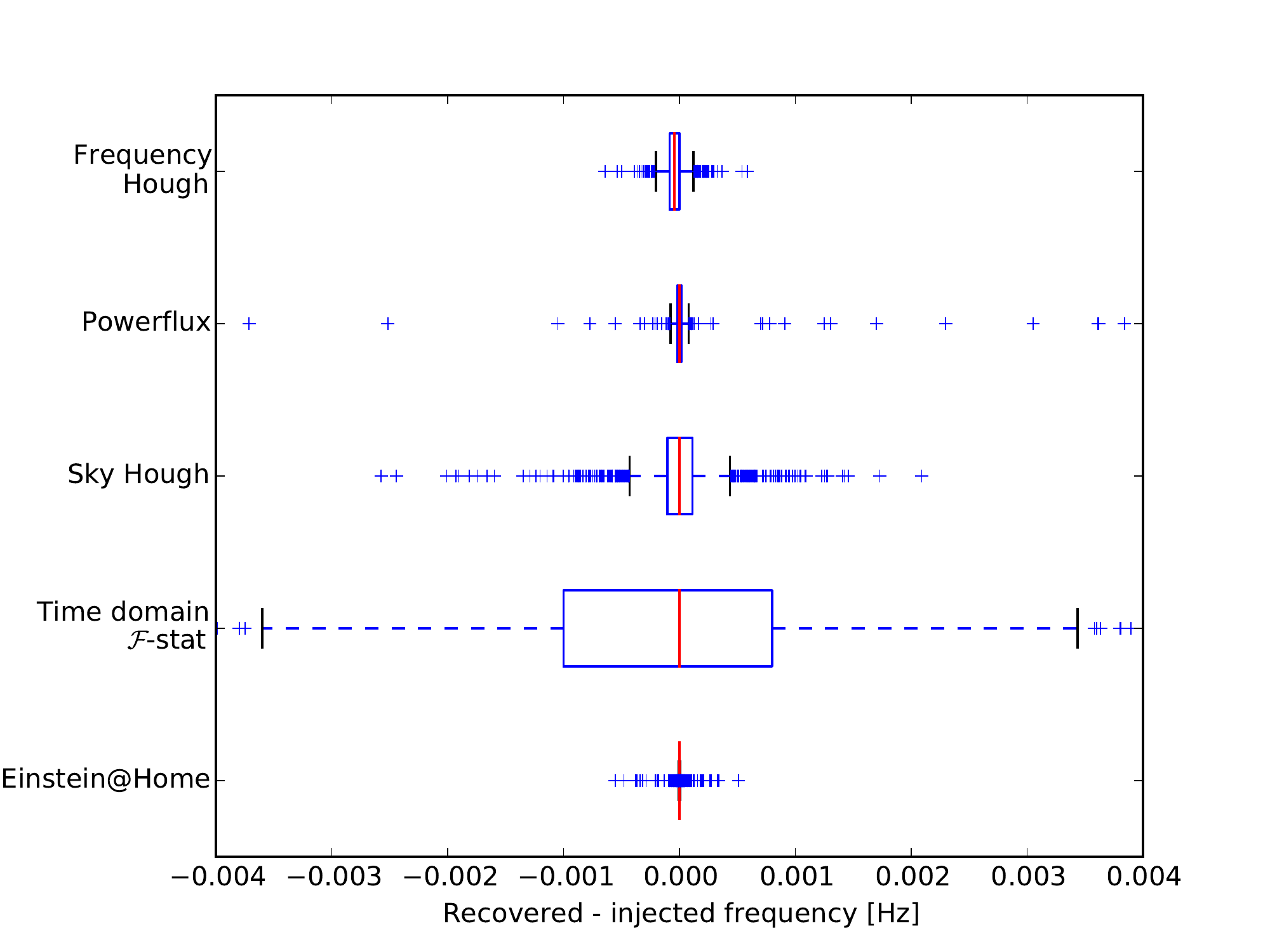}
  \includegraphics[width=3.2in]{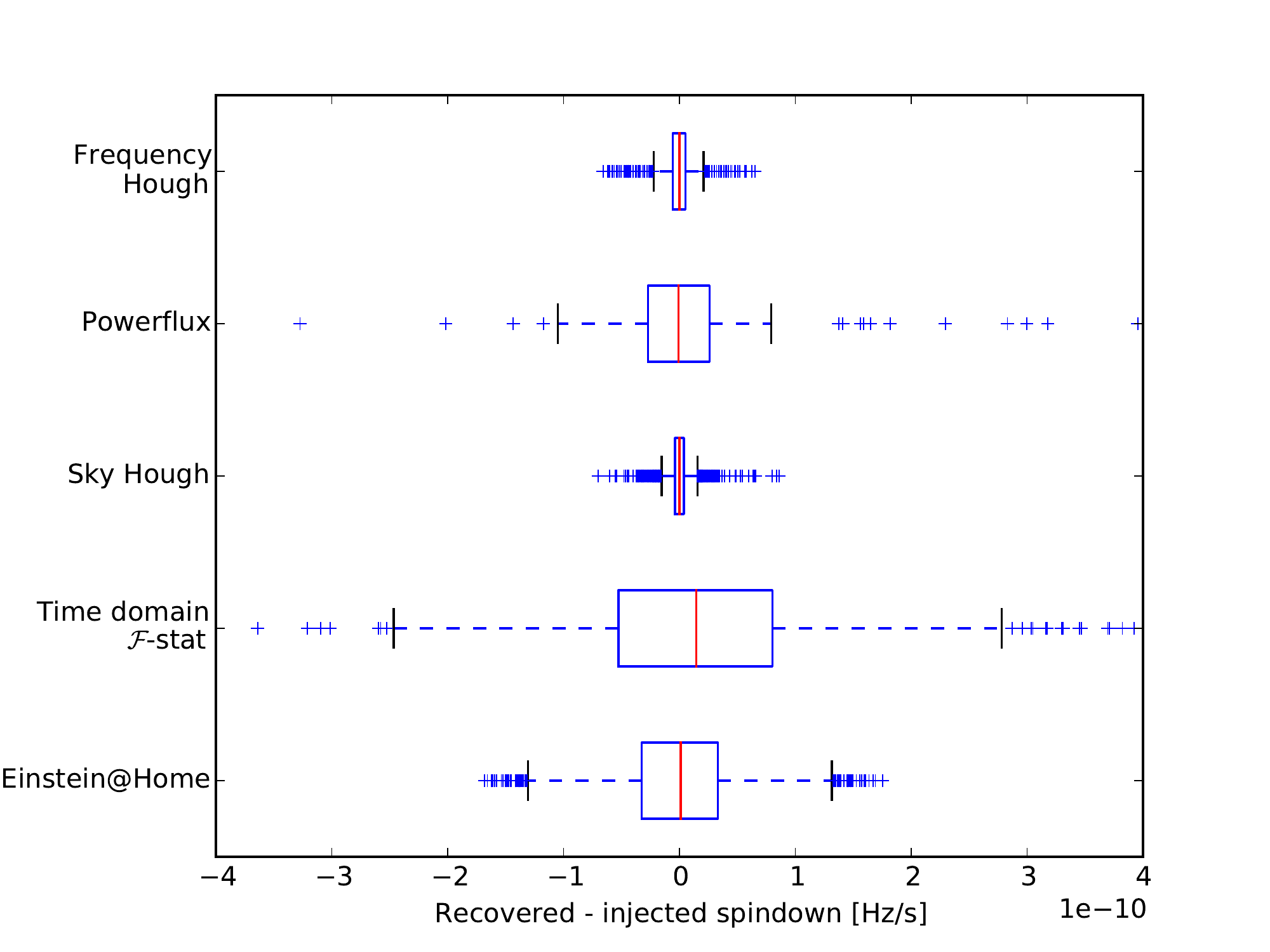}
  \includegraphics[width=3.2in]{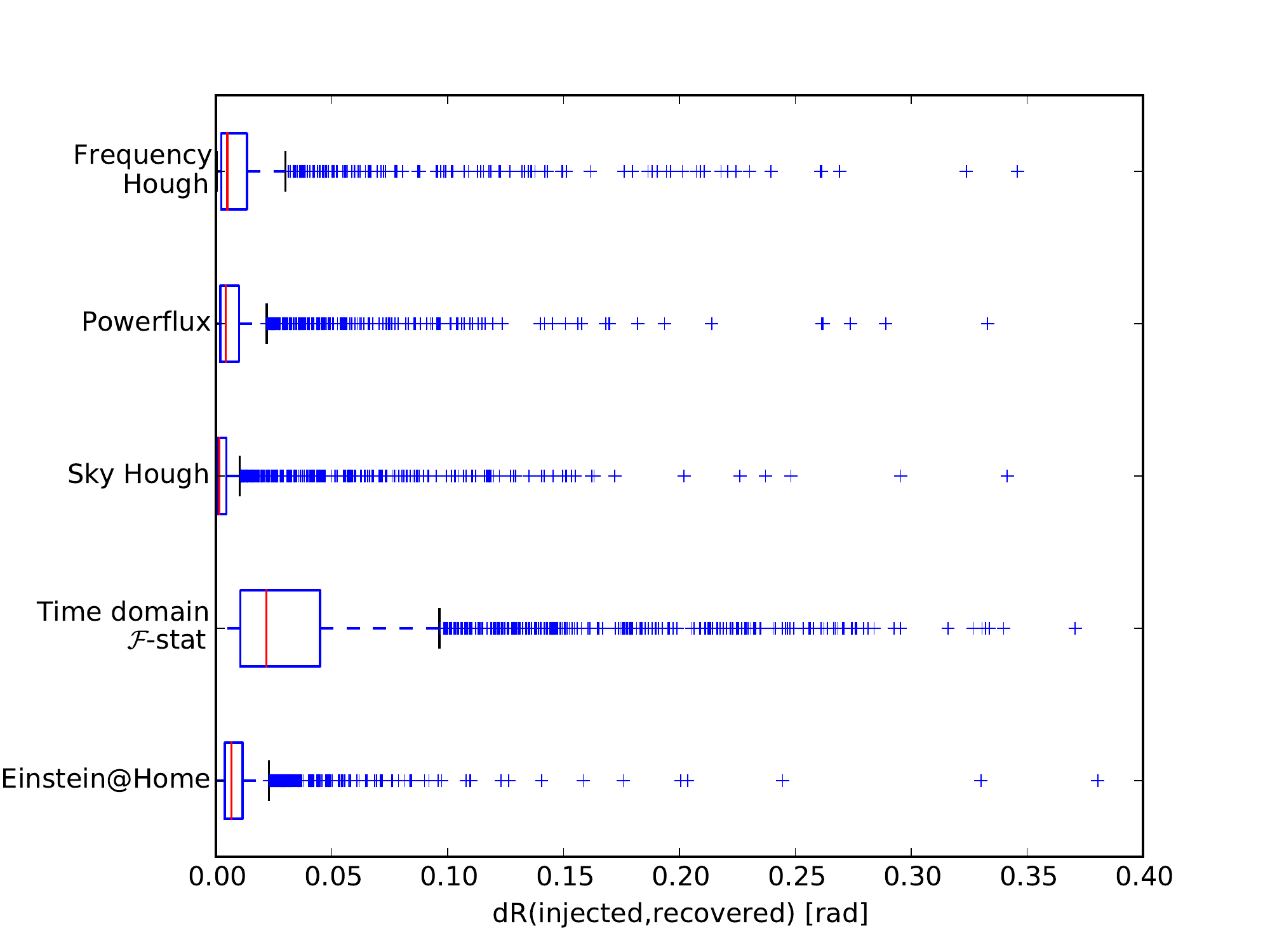}\\

\caption{\label{fig:distances_merged_minf0} The distance between the signal and the recovered candidate, in frequency, spindown and sky position, when the candidate with the closest frequency to the signal is chosen. The red line is the median. The blue box begins and ends at the first and third quartile respectively. The vertical black bars (whiskers) extend 1.5 times the inner quartile range outside the blue box. The blue crosses are candidates outside this range.}
\end{figure}

\begin{figure}[htb!]
  \includegraphics[width=3.2in]{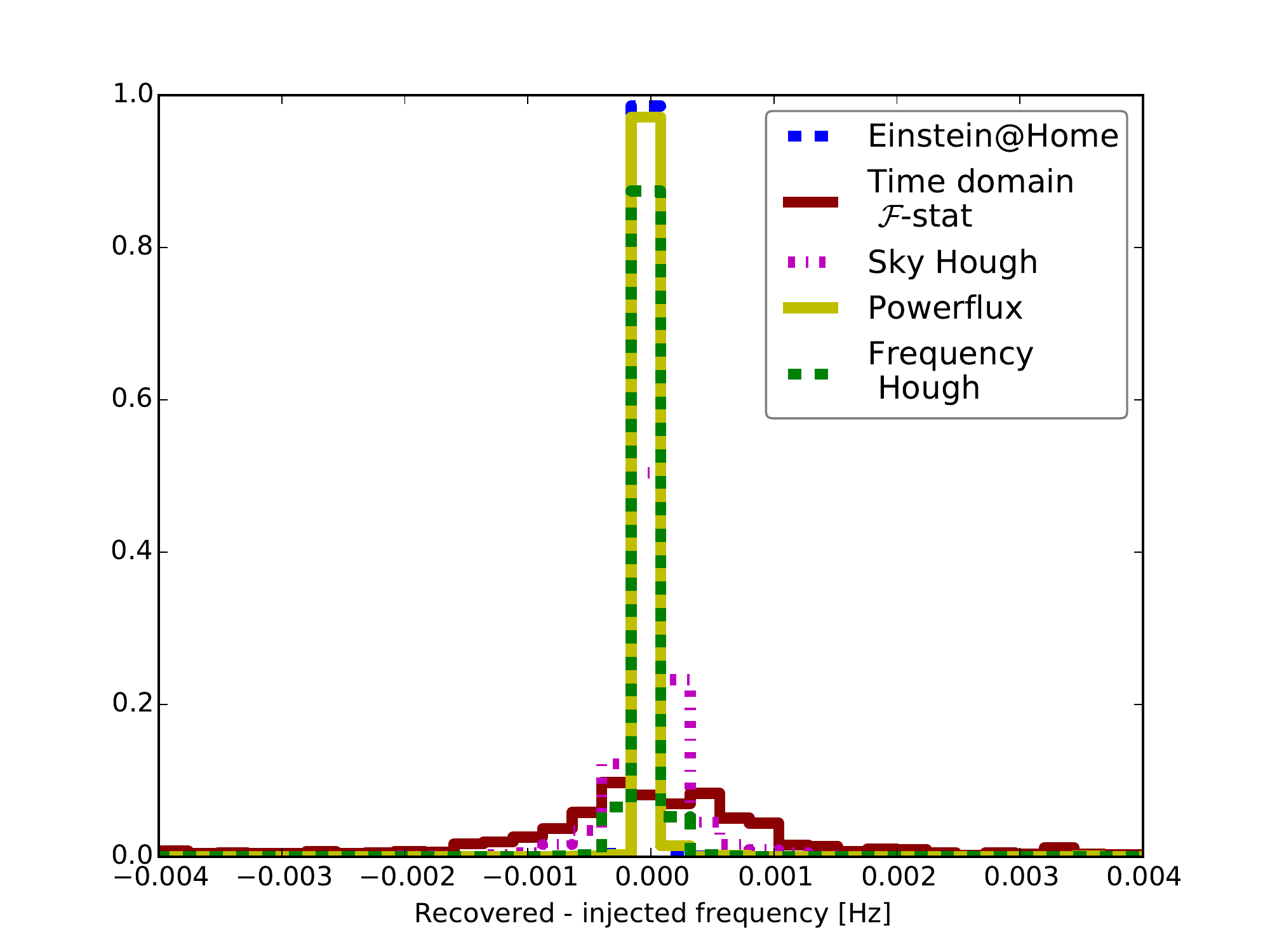}\\
  \includegraphics[width=3.2in]{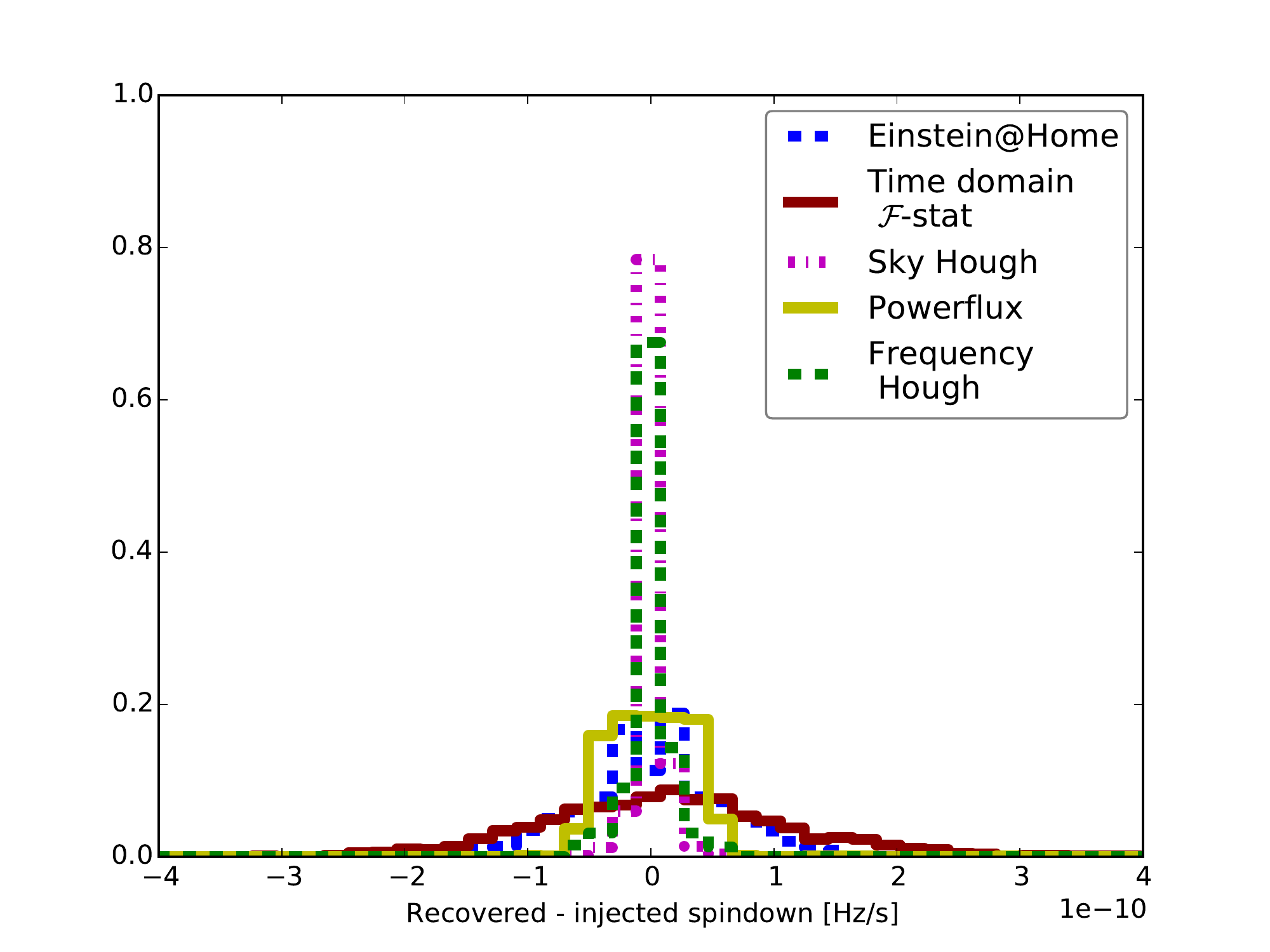}\\
  \includegraphics[width=3.2in]{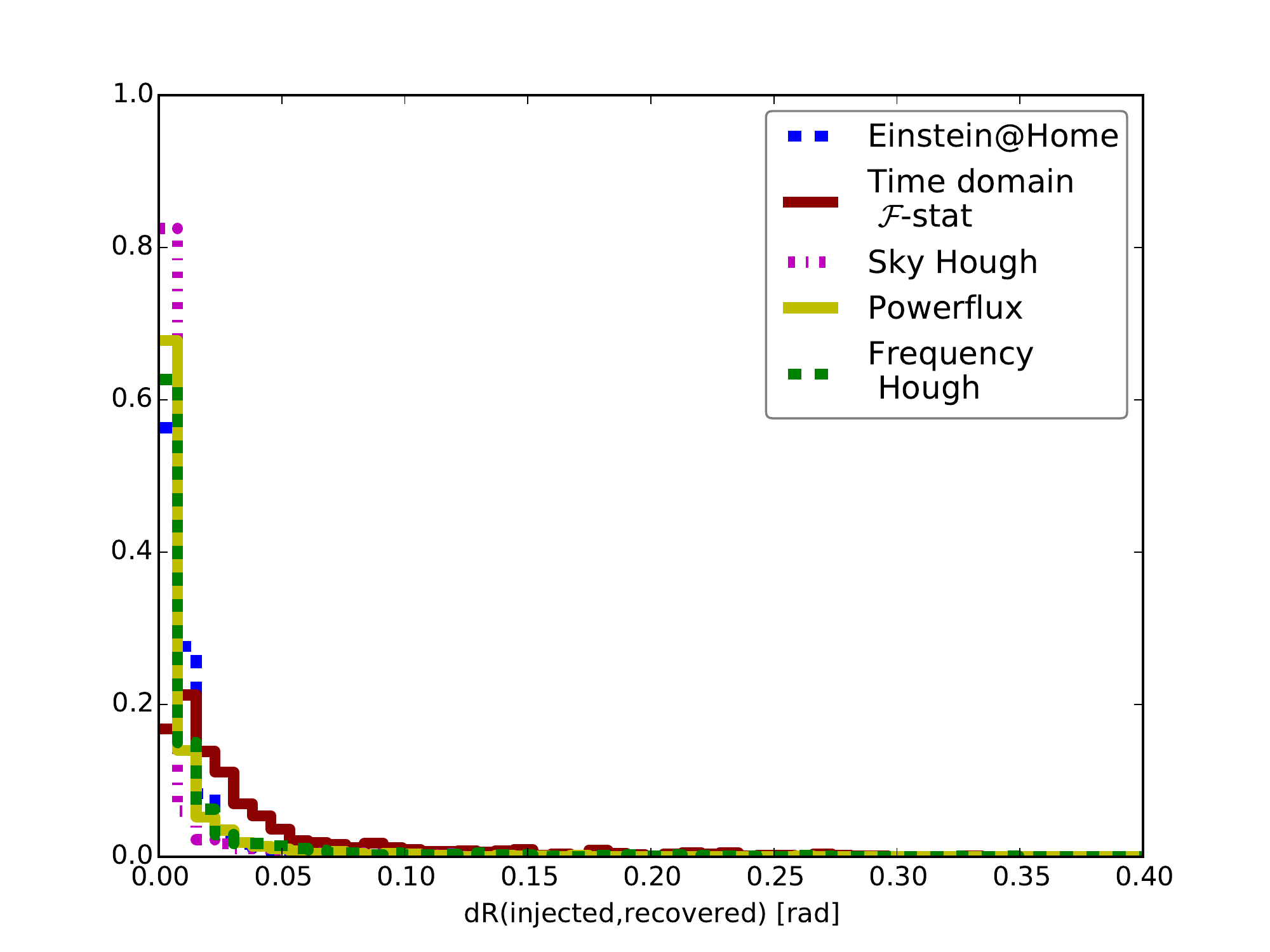}\\

\caption{\label{fig:distances_merged_minf0_hist} The distance between the signal and the recovered candidate, in frequency, spindown and sky position, when the candidate with the closest frequency to the signal chosen. This is an alternative representation of the same data as in Figure \ref{fig:distances_merged_minf0}.}
\end{figure}

Figures \ref{fig:distances_merged_minf0} and \ref{fig:distances_merged_minf0_hist} compare the distance between the signal and recovered candidate, now choosing the nearest candidate in frequency for Powerflux, Frequency Hough and Einstein@Home. The spread of the Powerflux results has decreased significantly, with the furthest outliers at $7\ee{-3}$\,Hz, $1.5\ee{-9}$\,Hz/s or dR 0.5 from the signal, and less than 1\% of candidates outside the boundaries in any dimension. Powerflux has demonstrated, in the search of S6 data \cite{PF_3}, that they are able to perform refinement searches on all candidates above threshold.

The Frequency Hough results are unchanged, as only a handful of signals result in more than one detection candidate. The inner quartile of the Einstein@Home distribution has changed. However, the presence of candidates at the edge of the refinement region shows that the parameter space can not be reduced without losing detection efficiency.\\

\section{Conclusion}

We have considered five search pipelines currently performing blind all-sky searches for continuous waves from isolated NSs in advanced detector data. An overview of each pipeline has been presented with regards to the semi-coherent method, noise handling, computing cost and the hierarchical refinement procedure. 

To compare the methods, an MDC was performed with $\sim 3000$ simulated signals. Each pipeline has presented the recovered signal candidates, from which the detection efficiency has been calculated. These results were used to compare the pipelines in quiet and noisy data, and to check for dependence on signal frequency and frequency derivatives. 

The search methods used by each pipeline make different tradeoffs in sensitivity vs.\ robustness against deviations from the assumed signal model. In this MDC, the detection efficiency is measured for a strictly continuous and phase coherent wave signal. This serves as a first benchmark in the comparison of these five all-sky pipelines. For a comprehensive comparison, the MDC must be extended to include signals that deviate from this signal model.
 
The precision with which the pipelines can be compared is restricted by the dependence of detection efficiency on the observing time of the data and the search configuration, which changes depending on the data available and the parameter space covered. Pipelines are also developing improvements, which will change the detection efficiency of future searches.

With these caveats in mind, we compare the detection efficiency of the pipelines for a standard CW signal. We find similar performance among the Sky Hough, Powerflux and Frequency Hough searches. The detection efficiency for these signals is lower for the Time domain $\mathcal{F}$-statistic search. The Einstein@Home search achieved comparable detection efficency to the other pipelines for signals that are a factor of two weaker, for frequencies below 1000\,Hz. The different noise handling approaches left the Sky Hough and Time domain $\mathcal{F}$-statistic efficiencies unchanged in the presence of known noise lines, while the Einstein@Home and Powerflux searches lost efficiency. The apparent dependence of detection efficiency on signal frequency for Einstein@Home and Powerflux is understood. There is no measured dependence on spindown for any search method.

Despite not explicity searching over second order spindown, the detection efficiency is unaffected by signals with non-zero second order spindown. This has been verified for $\ddot{f} < \ddot{f}_\mathrm{critical}$. Assuming a standard NS model, $\ddot{f}$ is not expected to exceed $\ddot{f}_\mathrm{critical}$ for the Sky Hough or Frequency Hough searches. The $\ddot{f}$ may exceed $\ddot{f}_\mathrm{critical}$ for the other searches. The impact of this has not been examined.

This study is a first step towards a quantative comparison of the different pipelines. Future studies are needed that include signals deviating from the standard CW model to understand and highlight the benefits of each pipeline.

\acknowledgments

The authors would like to thank Erin Macdonald for starting the setup of the MDC, and the LIGO and Virgo CW group for useful insights and discussions. The authors gratefully acknowlege support from the following grants. The Einstein@Home and Powerflux teams acknowledge the National Science Foundation grants NSF PHY 1104902 and NSF PHY 1505932, respectively. Time domain $\mathcal{F}$-statistic is supported by the National Science Centre of Poland grant UMO-2014/14/M/ST9/00707. Sky Hough is supported by the Spanish Ministerio de Econom\'ia y Competitividad (Grants No. FPA2013-41042-P,  CSD2009-00064, FPA2015-69815-REDT,  and FPA2015-68783-REDT), European Union FEDER funds, Conselleria d'Economia i Competitivitat del Govern de les Illes Balears and ``Fons Social Europeu''. M. P.\ is funded by the Science and Technology Facilities Council under Grant No.\ ST/L000946/1. The authors gratefully acknowledge the Italian Istituto Nazionale di Fisica Nucleare. This paper was assigned LIGO document number P1600128.

\appendix
\section{Comparing results with known and blind injections}
\label{app:stage3_stage4}
\begin{figure}[htb!]
  \includegraphics[width=3.5in]{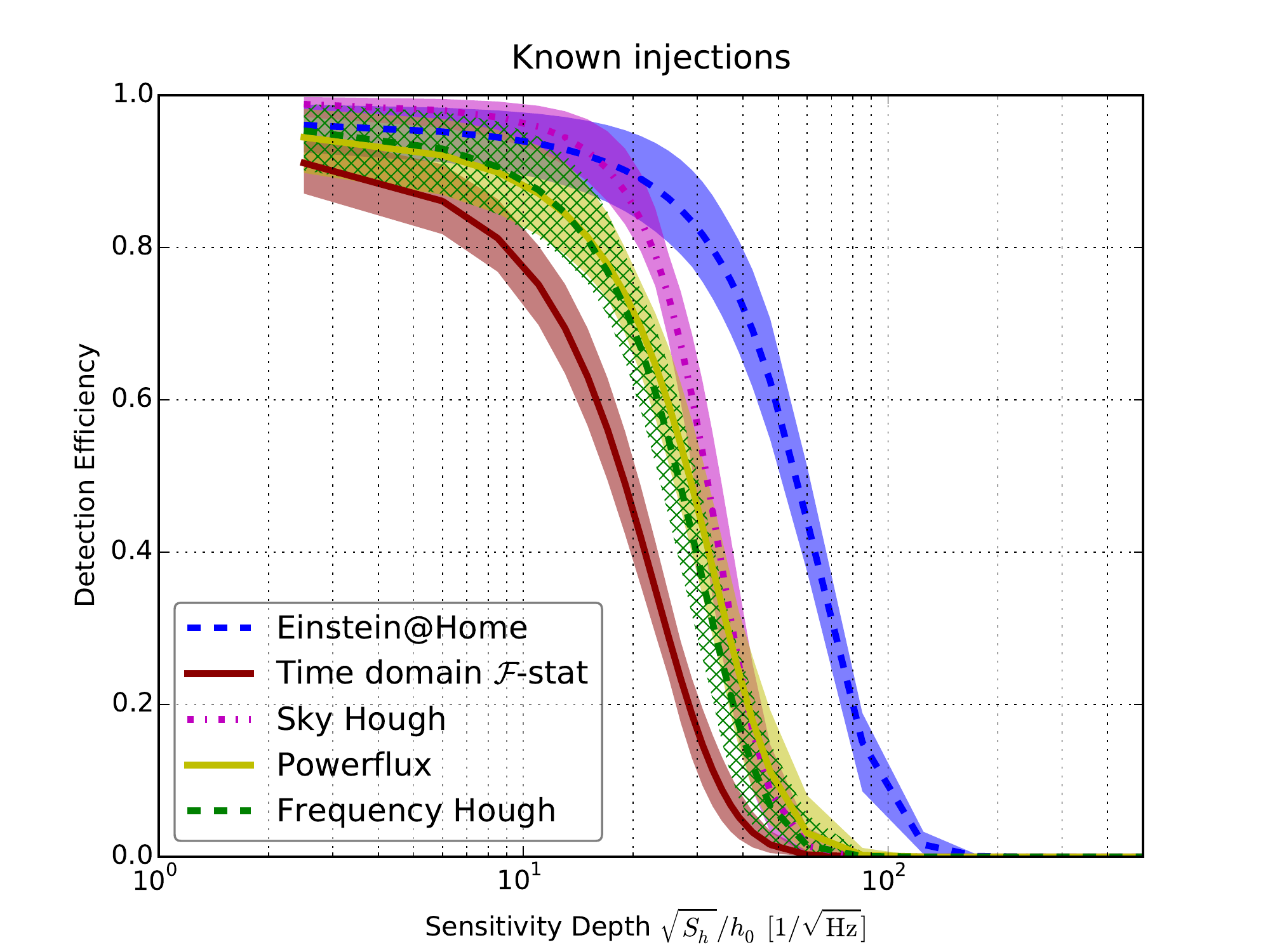}\\
 \includegraphics[width=3.5in]{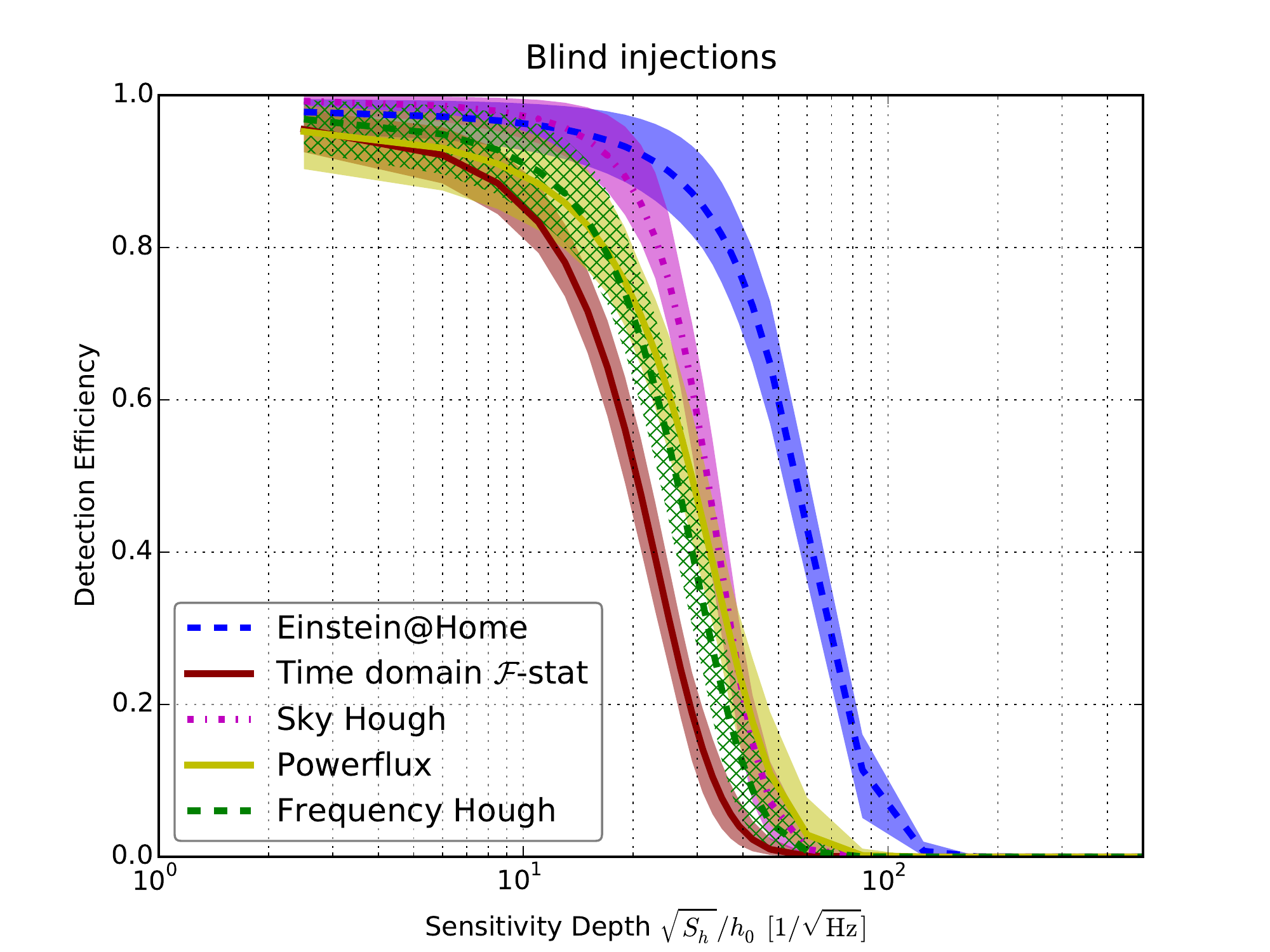}%
\caption{\label{fig:eff_stage3_stage4} Detection efficiency measured with known injections (top) and blind injections (bottom).}
\end{figure}

Figure \ref{fig:eff_stage3_stage4} shows that each search measures the same detection efficiency with injections where the injection parameters are known, and blind injections where the parameters are unknown. Therefore, we are able to combine the results from both injection sets. This reduces the statistical uncertainty on the measured detection efficiency.


\begin{thebibliography}{}

\bibitem{LIGO_Virgo_1} The LIGO Scientific Collaboration, ``LIGO: the Laser Interferometer Gravitational-Wave Observatory'', \textit{Rept. Prog. Phys} 72 (2009) 076901
\bibitem{LIGO_Virgo_2} The LIGO Scientific Collaboration and the Virgo Collaboration, ``Characterization of the LIGO detectors during their sixth science run'', \textit{Class. Quant. Grav.} 32 (2015) 115012
\bibitem{LIGO_Virgo_3} The Virgo Collaboration, ``Status of Virgo'', \textit{Class. Quant. Grav.} 25 (2008) 114045
\bibitem{EaH_S6} The LIGO Scientific Collaboration and the Virgo Collaboration, ``Results of an all-sky Einstein@Home search for continuous gravitational waves'', in preparation (2016)

\bibitem{PF_2} The LIGO Scientific Collaboration and the Virgo Collaboration, ``All-sky search for periodic gravitational waves in the full S5 LIGO data'', \textit{Phys. Rev. D} 85 (2012) 022001
\bibitem{FH_3} The LIGO Scientific Collaboration and the Virgo Collaboration, ``First low frequency all-sky search for continuous gravitational wave signals'', \textit{Phys. Rev. D} 93 (2016) 042007
\bibitem{TD_1} The LIGO Scientific Collaboration and the Virgo Collaboration, ``Implementation of an F-statistic all-sky search for continuous gravitational waves in Virgo VSR1 data'',  \textit{Class. Quantum Grav.} 31 (2014) 165014
\bibitem{SH_2} The LIGO Scientific Collaboration and the Virgo Collaboration, ``Application of a Hough search for continuous gravitational waves on data from the fifth LIGO science run'', \textit{Class. Quantum Grav.} 31 (2014) 085014 


\bibitem{Fstat} P. Jaranowski, A. Krolak and B. F. Schutz, ``Data analysis of gravitational-wave signals from spinning neutron stars: The signal and its detection'', \textit{Phys. Rev. D} 58 (1998) 063001
\bibitem{Owen_f2} B. Owen, ``How to adapt broad-band gravitational-wave searches for $r$-modes'', \textit{Phys. Rev. D} 82 (2010) 104002
\bibitem{VirgoCompCost} S. Frasca, P. Astone and C. Palomba, ``Evaluation of sensitivity and computing power for the Virgo hierarchical search for periodic sources'', \textit{Class. Quantum Grav.} 22 (2005) S1013
\bibitem{Brady} P. R. Brady and T. Creighton, ``Searching for periodic sources with LIGO. II. Hierarchical searches'', \textit{Phys. Rev. D} 61 (2000) 082001

\bibitem{SH_PF_1} The LIGO Scientific Collaboration, ``All-sky search for periodic gravitational waves in LIGO S4 data'', \textit{Phys. Rev. D} 77 (2008) 022001
\bibitem{PF_3} The LIGO Scientific Collaboration and the Virgo Collaboration, ``Comprehensive All-sky Search for Periodic Gravitational Waves in the Sixth Science Run LIGO Data'', LIGO Document P1500219
\bibitem{PF_LC_1} V. Dergachev, ``On blind searches for noise dominated signals: a loosely coherent approach'', \textit{Class. Quantum Grav.} 27 (2010) 205017

\bibitem{TD_sky} P. Astone, K. M. Borkowski, P. Jaranowski, M. Pi{\c e}tka and A. Kr\'olak, ``Data analysis of gravitational-wave signals from spinning neutron stars. V. A narrow-band all-sky search'', \textit{Phys. Rev. D} 82 (2010) 022005

\bibitem{EaH_url} einsteinathome.org
\bibitem{GCT_method} H. J. Pletsch and B. Allen, ``Exploiting Large-Scale Correlations to Detect Continuous Gravitational Waves'', \textit{Phys. Rev. Lett.} 103 (2009) 181102
\bibitem{BSGL} Keitel et al., ``Search for continuous gravitational waves: Improving robustness versus instrumental artifacts'', \textit{Phys. Rev. D} 89 (2014) 064023
\bibitem{EaH_S6_FU} M. A. Papa et al., ``Results of an all-sky Einstein@Home search for continuous gravitation waves with a new hierarchical approach'', in preparation (2016)

\bibitem{FH_1} Antonucci et al., ``Detection of periodic gravitational wave sources by Hough transform in the f versus f(.) plane'', \textit{Class. Quantum Grav.} 25 (2008) 184015
\bibitem{FH_2} P. Astone, A. Colla, S. D’Antonio, S. Frasca, and C. Palomba, ``Method for all-sky searches of continuous gravitational wave signals using the frequency-Hough transform'', \textit{Phys. Rev. D} 90 (2014) 042002


\bibitem{S6data}  The LIGO Scientific Collaboration and the Virgo Collaboration, ``Characterization of the LIGO detectors during their sixth science run'', \textit{Class. Quantum Grav.} 32 (2015) 115012
\bibitem{Lalsuite} L. D. A. S. W. Group, ``Lalsuite''
\bibitem{Wette_f2} K. Wette et al., ``Searching for gravitational waves from Cassiopeia A with LIGO'', \textit{Class. Quantum Grav.} 25 (2008) 235011
\bibitem{Cristiano_f2} C. Palomba, ``Pulsars ellipticity revised'', \textit{A\&A} 354 (2000) 163

\bibitem{AllenK} M. P. Allen and J. E. Horvath, ``Glitches, torque evolution and the dynamics of young pulsars'', \textit{MNRAS} 287 (1997)


\bibitem{Schaltev} M. Shaltev, P. Leaci, M. A. Papa and R. Prix, ``Fully coherent follow-up of continuous gravitational-wave candidates: an application to Einstein@Home results'', \textit{Phys. Rev. D} 89 (2014) 124030


\bibitem{SH_new} M. Oliver et al., ``Improvement and characterization of the Hough search for all-sky continuous gravitational wave signals'', in preparation (2016)

\bibitem{Reinhard} R. Prix and M. Shaltev, ``Search for Continuous Gravitational Waves: Optimal StackSlide method at fixed computing cost'', \textit{Phys.Rev. D} 85 (2012) 084010
\bibitem{EaH_S5} The LIGO Scientific Collaboration and the Virgo Collaboration, ``Einstein@Home all-sky search for periodic gravitational waves in LIGO S5 data'', \textit{Phys. Rev. D} 87 (2013) 042001




\end{thebibliography}
\end{document}